\documentclass[fleqn,usenatbib]{mnras}

\usepackage{mathptmx}
\usepackage{hyperref}

\usepackage[T1]{fontenc}
\usepackage{ae,aecompl}

\usepackage{graphicx}	
\usepackage{amsmath}	
\usepackage{amssymb}	

\usepackage{txfonts}
\usepackage{xcolor}
\usepackage{soul}

\usepackage[export]{adjustbox}

\title[Machine Learning \& Multifrequency Astrophysical Data]{Machine Learning applied to Multifrequency Data in Astrophysics: Blazar Classification}

\author[B. Arsioli \& P. Dedin]{B. Arsioli ,$^{1,2}$\thanks{E-mail: arsioli@ifi.unicamp.br \& bruno.arsioli@gmail.com}
and P. Dedin,$^{1}$\thanks{E-mail: dedin@ifi.unicamp.br}
\\
$^{1}$Gleb Wataghin Physics Institute, Unicamp, Rua S. Buarque de Holanda 777, 13083-859, Campinas, BR\\
$^{2}$ICRANet, P.zza della Repubblica 10, I-65122, Pescara, IT
}

\date{Accepted XXX. Received yyy; in original form ZZZ}

\pubyear{2020}

\begin{document}
\label{firstpage}
\pagerange{\pageref{firstpage}--\pageref{lastpage}}
\maketitle

\begin{abstract}
 
The study of machine learning (ML) techniques for the autonomous classification of astrophysical sources is of great interest, and we explore its applications in the context of a multifrequency data-frame. We test the use of supervised ML to classify blazars according to its synchrotron peak frequency, either lower or higher than 10$^{15}$Hz. We select a sample with 4178 blazars labelled as 1279 high synchrotron peak (HSP: $\rm \nu$-peak\,>\,10$^{15}$Hz) and 2899 low synchrotron peak (LSP: $\rm \nu$-peak\,<\,10$^{15}$Hz). A set of multifrequency features were defined to represent each source, that includes spectral slopes ($\alpha_{\nu_1, \nu_2}$) between the radio, infra-red, optical, and X-ray bands, also considering IR colours. We describe the optimisation of five ML classification algorithms that classify blazars into LSP or HSP: Random Forests (RF), Support Vector Machine (SVM), K-Nearest Neighbours (KNN), Gaussian Naive Bayes (GNB) and the Ludwig auto-ML framework. In our particular case, the SVM algorithm had the best performance, reaching 93\% of balanced-accuracy. A joint-feature permutation test revealed that the spectral slopes alpha-radio-IR and alpha-radio-optical are the most relevant for the ML modelling, followed by the IR colours. This work shows that ML algorithms can distinguish multifrequency spectral characteristics and handle the classification of blazars into LSPs and HSPs. It is a hint for the potential use of ML for the autonomous determination of broadband spectral parameters (as the synchrotron $\nu$-peak), or even to search for new blazars in all-sky databases.
\end{abstract}

\begin{keywords}
methods: data analysis --  methods: statistical --  galaxies: active -- radiation mechanisms: non-thermal
\end{keywords}




\section{Introduction}

The autonomous classification of astrophysical sources is a challenge faced by modern astronomy, given the growing scale of its databases. High sensitivity detectors have opened a window for deep-sky searches able to reach distant and faint sources that are far too numerous to handle case-by-case. The study of efficient classification and selection schemes via algorithmic modelling \citep{breiman2001} has become vital to explore big datasets and to help deliver refined scientific products. Consider, for example, the upcoming deep radio survey -i.e., the Evolutionary Map of the Universe \citep[EMU][]{ASKAP-EMU-Hopkins2015}- that is one of the leading science objectives from the Australian Square Kilometre Array Pathfinder (ASKAP). The ASKAP-EMU data-base will be crucial to search for early AGNs \citep{SMBH-ASKAP-Amarantidis2019} and is expected to identify nearly 75 million point sources down to 10 $\mu$Jy, to compare to $\sim$2.5 million from NVSS \citep{Condon1998}, our deepest radio survey up to date. In this context, the classification of astrophysical objects will widely profit from machine learning techniques, which are ready to implement and open to fine-tuning optimisation. 

We use multifrequency spectral data (from radio up to X-rays) to train, test, and compare several machine learning (ML) models applied to the classification of blazars according to the energy associated with its synchrotron peak. We show that ML codes can classify blazars based on their multifrequency spectral features, setting the ground for future works that aim to search for new blazars within multifrequency databases. 

Blazars are a rare type of active galactic nuclei (AGN) known for its multifrequency emission over the entire electromagnetic spectrum and characterised by fast spectral variability. The central engines from AGNs are powered by matter accretion into supermassive black holes, which produce jets of relativistic charged particles \citep{Padovani2017}. When those jets point close to our line of sight, the sources are called blazars, and the observer perceives a bright object due to relativistic boosting effects. The propagation of relativistic particles along the magnetised jets originates a non-thermal component which can cover many decades in energy, from radio up to TeV gamma-rays. 

\begin{figure}
    \centering
    \includegraphics[width=1.0\linewidth]{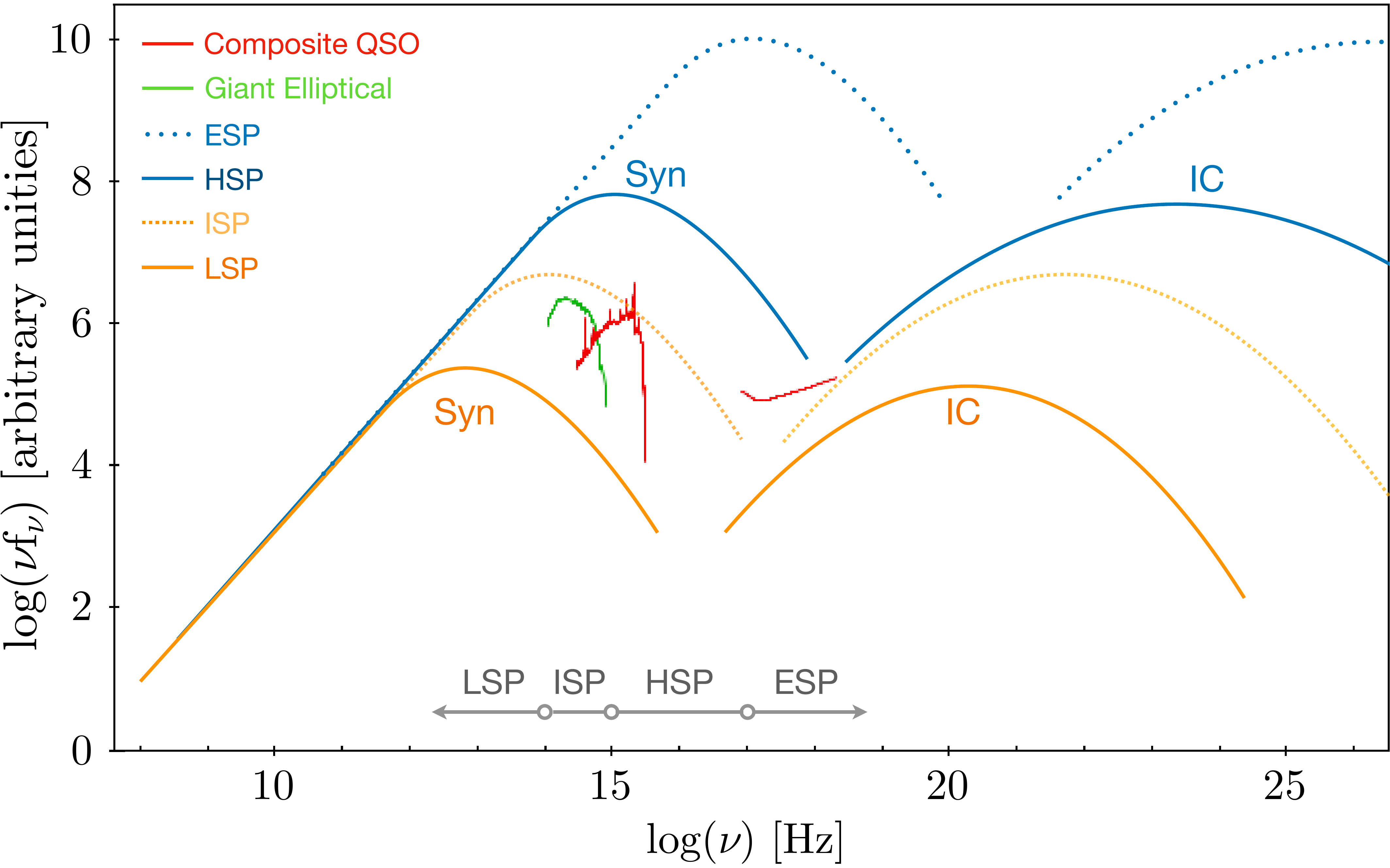}
    \caption{A qualitative SED showing the blazar types according to an energy-based classification: low, intermediate, high, and extreme synchrotron peak [with $\nu$ in Hz]: LSP for log($\rm \nu_{peak}$)\,<\,14 (orange line), ISP for 14\,<\,log($\rm \nu_{peak}$)\,<\,15 (orange dashed line), HSP for 15\,<\,log($\rm \nu_{peak}$)\,<\,17 (blue line), and ESP for log($\nu_{peak}$)\,>\,17 (blue dashed line), highlighting the synchrotron (Syn) and Inverse Compton (IC) humps. Besides, we show spectral templates that represent the thermal emission from Giant Elliptical galaxies 
   \citep[green lines,][]{Elliptical-Template-Mannucci2001}, and the expected Composite Quasar Spectrum from accretion (also known as `blue bump') with its associated X-ray emission \citep[red lines,][]{Composite-QSO-2001}. The relative intensity between thermal and non-thermal components vary for each source, and a mixture is always at play in blazar's SED.}
    \label{fig:blazar-classes}
\end{figure}

The presence of two non-thermal humps in the spectral energy distribution (SED) of blazars (Fig.~\ref{fig:blazar-classes}) originates from synchrotron and inverse Compton (IC) emission processes due to relativistic electrons moving through the jet's magnetic field lines \citep{Giommi2015,giommi2012}. Moreover, the thermal emission from the host galaxy and the accretion disk are both crucial elements to describe the overall spectrum, and a mixture of thermal and non-thermal components are at play \citep{Urry-Padovani1995}, hindering autonomous identification and classification efforts.    

A variety of selection criteria are proposed in the literature to search and identify new blazars \citep{Arsioli2015,Chang2017,DiMauro2014,DAbrusco2012,Massaro2015} and are mostly based on the definition of qualitative boundaries in colour-colour planes, or by defining selection rules based on multifrequency spectral slopes. While one tries to simplify the selection rules and make it more tangible to comprehend, there is usually a loss of efficiency involved, meaning the final selection gets compromised with spurious sources and human intervention (as case-by-case evaluation) is necessary to clean the final sample.

ML algorithms are versatile in dealing with selection and classification problems in a multi-parameter space, where numerous input features are available \citep{DataMine-ML-Astronomy-2010,Borne-DataMine-Astro-2009}. A good example of challenging astrophysical label problem is to disentangle `stars' from `galaxies' in optical observations. Common non-ML strategies used to handle the star-galaxy separation task involve colour-colour and proper motion selection schemes \citep{HST-counts-RogierWindhorst-2011}. Alternatively, \cite{QSO-class-ML-2019} has shown that multifrequency data from optical and IR, together with ML techniques, can largely improve the accuracy to label stars, galaxies and quasars when compared to single-band classification methods.

In the case of blazars, the classification follows two main roads. One only based on the characteristics of the optical spectra, and another based on the energy (frequency) associated with the synchrotron peak component. For the optical classification, blazars can be divided into BL Lacs (when the optical spectra is dominated by the jet's non-thermal component, therefore featureless), Flat Spectra Radio Quasar (FSRQ, when the optical spectra is dominated by thermal components from the AGN's core, showing the blue bump with strong and broad emission lines), and when the optical features are not clear (with weak or diluted emission lines within a relatively strong non-thermal component) the term uncertain/transitional blazar is used \citep{Massaro2009,Massaro2015}. 

For the energy-based classification of blazars, the definition of families considers the synchrotron peak frequency (Fig. \ref{fig:blazar-classes}). Accordingly, blazars are called as low, intermediate, high, and extreme synchrotron peak sources: LSP for $ \rm \nu^{syn}_{peak}$\,<\,$\rm 10^{14}$\,Hz, ISP for $\rm 10^{14}$\,<\,$\rm \nu^{syn}_{peak}$\,<\,$\rm 10^{15}$\,Hz, HSP for $\rm 10^{15}$\,<\,$\rm \nu^{syn}_{peak}$\,<\,$\rm 10^{17}$\,Hz, and ESP for $\rm \nu^{syn}_{peak}$\,>\,$\rm 10^{17}$\,Hz \citep{padgio95,BlazarSED,ExtremeBlazarsGhisellini1999}.

Both classification schemes demand a substantial human intervention to treat objects case-by-case, either to inspect optical spectral data to decide in between BL Lacs, FSRQ and Uncertain types, or to fit the non-thermal component and measure the synchrotron peak frequency. For the case of optical classification, \cite{BCU-ML-Class-2019} attempts to use ML models to improve the labelling of uncertain blazars (BZUs) from the 3LAC catalogue \citep{3LAC}. Their ML modelling predicts an FSRQ:BL Lac fraction of 1:3 for the 3LAC-BZUs. Other works investigate the same issue using a ML approach: \cite{Chiaro-Kovaevi2020,Lefaucheur2017,Yi2017,Chiaro2016,Massaro2016-fbratio}. 

Here we use an ML algorithmic approach to explore aspects of the energy-based classification of blazars, and train models to perform the task of labelling HSP and LSP sources ($\rm \nu^{syn}_{peak}$>$\rm 10^{15}$Hz and $ \rm \nu^{syn}_{peak}$<$\rm 10^{15}$Hz, respectively) only based on multifrequency information (fluxes and derived spectral slopes). Our long term goal, however, is to inspect and identify strategies to search for new blazars within large-scale astrophysical databases. 


\section{Methodology}
There are currently two main blazar catalogues, namely the 5BZcat \citep{Massaro2015} with 3561 objects, and the 3HSP \citep{Chang2019} with 2013 objects (Note: 657 3HSPs are already listed in 5BZcat). The 5BZcat and 3HSP sum a total of 4915 unique sources which are either confirmed blazars, or blazar-candidates\footnote{The term `confirmed blazar' refers to optically identified sources; therefore, an optical spectrum is available. All 5BZcat sources are confirmed blazars. The term `blazar-candidate' refers to sources that lack optical identification. The 3HSP catalogue includes confirmed and blazar-candidates.} and are well characterised from radio to the X-ray band. Moreover, a large fraction has counterparts in the latest $\gamma$-ray catalogues, the 4FGL and 2BIGB \citep{4FGL,2BIGB}. In Section \ref{sec:selection}, we describe the use of high-accuracy labelled data from the 5BZcat and 3HSP catalogues to select a robust blazar sample, as well as the multifrequency cross-matching to build the ML data-frame. In Section \ref{sec:feature-selection}, we use multifrequency fluxes to define spectral slopes between channels and apply statistical tests to select features for training our ML models. 

In Section \ref{sec:MLalgos} the ML algorithms: RF, SVM, KNN, GNB, and Ludwig are briefly described, as well as the main evaluation metrics used in this work, and the re-sampling process used for the model's validation process. In Section \ref{sec:MLoptmization}, there is a description of the optimisation for the Random Forest (RF), Support Vector Machine (SVM), K-Nearest-Neighbours (KNN), and Gaussian Naive Bayes (GNB) algorithms \citep{Breiman2001-RF,Cortes1995-SVM-Network,KNN-Jacob-2004,NB-NaiveBayes-1998}. The optimisation is carried by scanning over the parameter space relevant for each algorithm using the SciKit learn libraries in Python \citep{scikit-learn}. We also test the performance of Ludwig, a deep-learning (auto-ML) tool released by the Uber team, but this time letting the tool auto-adjust its parameters (without user fine-tuning) to compare to our optimised models.

In Section \ref{sec:tunning}, the level of balanced and absolute accuracy, precision and recall are used to evaluate and compare the performance of all ML models. We take into account of statistical fluctuations and report on mean output values as derived from hundreds of random re-samples. We track the metrics standard deviation as a direct way to evaluate the stability and robustness of each algorithm. In Section \ref{sec:feature-importance}, we evaluate the importance of each input feature for optimised models, applying the so-called permutation test (a column associated to a given feature is shuffled, and the impact over the model accuracy is measured). The correlation between features is taken into account, and joint-feature permutation tests (shuffle of multiple features at once) are also described. In Section \ref{sec:miss-rate} we evaluate the performance of the optimised algorithms by tracking the misclassication-rate, which consists in counting the number of times an object is misclassified (by at least one algorithm) when considering 1000 re-samples. This way, we can identify and flag the most ambiguous cases.


\section{The ML Blazar sample}
\label{sec:selection}

The need to combine multifrequency databases is a known challenge in astrophysics, and currently, there is no autonomous solution for such a problem. To fuse databases is one of the most significant issues in the fields of Machine Learning and Artificial Intelligence as of today, and the building of data-frames is a fundamental step where substantial human effort is indeed needed (involving data handling, querying, cross-match, and fusion). There are ongoing and successful efforts to solve the data integration problem in astrophysics, and the Virtual Observatory \citep[VO][]{VO-2010} is a sound example. Alternatively, there is software in place as TopCat \citep{TopCat} that can handle big data-sets in astrophysics, offering numerous cross-match and visualisation tools.  

To build our data-frame we add a categorical attribute called `blazar-type' to flag sources either as HSP+ESP (flag 1) or LSP+ISP (flag 0). The 5BZcat is the largest catalogue of optically identified (confirmed) blazars but has no synchrotron peak ($\nu^{syn}_{peak}$) associated with its sources. The categorical labels available from 5BZcat rely on optical classification: BZB for BL Lacs, BZQ for FSRQs, and BZU for uncertain/transitional sources. To label 5BZcat sources as HSP or LSP, we use information from the 3HSP catalogue. The 3HSP \cite{Chang2019} is built over the search for HSP sources taking into account multifrequency selection rules together with a direct search within the 5BZcat and 3FHL \citep{Fermi3FHL2017} catalogues. Therefore, any object from 5BZcat that has no counterpart in 3HSP is considered as LSP (flag 0), and we label all 5BZcat-3HSP as HSP (flag 1). We use TopCat to cross-match the 5BZcat and 3HSP with 30'' radius (30 arc-seconds) and get 658 5BZcat sources classified as HSP+ESP (HSP: flag 1) and 2903 classified as LSP-ISP (LSP: flag 0).  

At this stage, a blazar sample only based on 5BZcat sources is relatively unbalanced, because there are approximately 4.41$\times$ more LSPs than HSPs. There are many possible strategies to remedy the unbalance. During the ML model training, some algorithms allow to increase the weight associated to the less frequent class (the HSPs) or even generate new HSPs based on population properties of the HSP subsample  \citep[with the SMOTE\footnote{SMOTE: Synthetic Minority Over-sampling Technique, is an algorithm that identifies the minority class and randomly select members of it, look into the k-Nearest Neighbours (n-neighbours=5) accounting for the available features, and produce new members of the minority class. The final sample becomes balanced to all classes.} technique,][]{SMOTE-2002}. In any case, the best approach to reduce the unbalance is to add extra real HSP sources to the sample.

The 5BZcat lists only 657 of the 2013 3HSP sources. Although the expected level of contamination for the 1356 3HSP-out-of-5BZcat is low \citep{Chang2019}, we only select cases that have a confirmed $\gamma$-ray counterpart from the 4FGL \citep{4FGL} or the 2BIGB \citep{2BIGB} catalogues. This selection translates into 627 additional HSPs: 482 that have a $\gamma$-ray counterpart in 4FGL and another 145 in 2BIGB. From those, 174 lack optical identification. 

We rely on the $\gamma$-ray detection as a proxy to assure the blazar nature of the additional HSPs, keeping a high degree of purity in the data-frame. Finally, we define the ML-Blazar sample as the sum of the entire 5BZcat (3561 sources) with the $\gamma$-ray detected 3HSP-out-of-5BZcat (627 sources). At this stage, the ML-Blazar sample has a total of 4188 sources, classified as 1285 HSPs and 2903 LSPs, which improves the 5BZcat LSP/HSP unbalance from 4.41$\times$ to a factor of 2.25$\times$. In the next section \ref{sec:preprocess}, there are additional multifrequency constrains, especially from missing data that force to reduce the ML-Blazar sample to 4178 sources (1279 HSPs and 2899 LSPs).

\subsection{Preprocessing multifrequency data}
\label{sec:preprocess}

The spectral energy distribution (SED) from 5BZcat and 3HSP blazars are well described along the radio to the X-ray band; therefore, the observed fluxes can become input attributes to train ML classification models. To build our multifrequency data-frame, we focus on the radio, IR, optical and X-ray fluxes, which have extensive available coverage. To search for archival spectral data and cross-match between catalogues we made use of several web portals and tools: Vizier \citep{Vizier}, Open Universe \citep{OpenUniverse}, the Sky-Explorer Tool from the Space Science Data Centre - Italian Space Agency (SSDC-ASI), and the TopCat software \citep{TopCat}. In the following paragraphs, we highlight the main procedures to collect and fuse multifrequency information for each band and feed our data-frame.

{\bf Radio.} We read the radio fluxes reported in the 5BZcat and 3HSP catalogues, which comprises measurements from five different surveys: FIRST (1.4 GHz), NVSS (1.4 GHz), SUMSS (0.843 GHz), PMN (4.8 GHz) and TAPMN (4.85 GHz) \citep{Helfand2015,Condon1998,Manch2003,PMN-radio,ATPMN-radio}. We keep track of the frequency $\nu_{radio}$ [Hz] associated to each measurement and use it for the calculation of broadband spectral slopes, following eq. \ref{eq:slope}. Within the ML-Blazar sample, 29 sources are not detected in any radio survey. Nevertheless, those sources constitute robust blazars, well-characterised and detected in IR, optical, UV, X-rays, and most of them (25) are also detected in $\gamma$-rays with Fermi-LAT (4FGL and 2BIGB). Considering that NVSS and SUMSS have surveyed the entire radio-sky, we include upper limit (UL) values for the radio fluxes (2.5 mJy and 4.15 mJy, respectively, considering the NVSS and SUMSS coverage)\footnote{Following \cite{Condon1998}, the completeness limit for NVSS is about 2.5 mJy (2.5$\times$10$^{-26}$ erg/cm2/s/Hz) at 1.4 GHz, and similar for SUMSS \citep{Manch2003} which converts to 4.15 mJy at 0.843 GHz.}. We add a categorical attribute to our data-frame called `radio-flag': `0' to flag sources already detected in radio, and `1' for the non-detections (with UL values). 

{\bf IR.} For the infra-red (IR) band, we rely on data from the Wide-field Infrared Survey Explorer and use the AllWISE survey catalogue \citep{Cutri2013}. The flux measurements done at the w1 (8.817E13 Hz), w2 (6.517E13 Hz) and w3 (2.498E13 Hz) channels are incorporated to the ML data-frame. We avoid the w4 (1.363E13 Hz) channel since it has the worst sensitivity. The ML-Blazar and WISE databases are cross-matched considering a 20'' radius, and only the best matches are kept. Eight sources showed no IR counterpart due to the presence of bright stars in their vicinity that hinders the IR detection. Those were eliminated from the ML data-frame, which now remains with 4180 sources. The WISE catalogue reports on magnitude values in Vega System unity and we follow the WISE team instructions\footnote{ \href{http://wise2.ipac.caltech.edu/docs/release/allsky/expsup/sec4_4h.html}{The WISE IPAC documentation}.} to convert magnitudes into flux density values [erg/cm$^2$/s], considering the case of an underlying point-sources with power-law spectra. 



{\bf Optical.} For the optical band, the ML-Blazar sample was cross-matched with four catalogues, including (in priority order): GAIA DR2, SDSS DR12, PanStars DR1, and Usno B.1 \citep{GAIA-DR2,SDSS-DR12,Chambers2016,Monet2003}, considering 3'' radius. We keep only the best match within each optical database (Table \ref{tab:optical-samples}) and concatenate all matches in a single table; that one has 13384 correspondences between the ML-Blazar sample with the four optical catalogues. All repeated sources were removed by running an internal cross-match with TopCat, keeping a single optical match for each ML-Blazar source, and considering the priority order for the optical catalogues. As a result, 4122 out of 4180 ML-Blazars (98.6\%) have optical information assigned. Given that the 5BZcat also reports the Rmag (4.454$\times$10$^{14}$Hz) for most sources, we could remedy 56 out of the 58 cases that were missing optical data. Finally, we keep 4178 sources in the data-frame, all assigned to an optical counterpart.

\begin{table}
	\centering
	\caption{Spectral information about the cross-match between the ML-Blazar sources with four optical catalogues (in top-down priority order).}
	\label{tab:optical-samples}
	\begin{tabular}{lccr} 
		\hline
		Catalog & Channel name & $\rm \nu \times 10^{-14}$ [Hz] & N matches \\
		\hline
		GAIA DR2     & G$_{rp}$ & 3.761  & 3840 \\ 
        SDSS DR12    & r        & 4.862  & 2073 \\
        PanStars DR1 & R        & 4.823  & 3523 \\ 
        Usno B.1     & R        & 4.454  & 3948 \\
		\hline
	\end{tabular}
\end{table}

{\bf X-ray.} To start collecting information for the X-ray band, we read the 1 keV fluxes (at $\nu_{1 keV}$=2.42$\times$10$^{17}$Hz) as reported in the 5BZcat and 3HSP catalogues. The ML-Blazar sample includes 2246 X-ray flux values taken from 5BZcat and 448 from the 3HSP catalogue. Some few cases having no X-ray information in 5BZcat now have available flux reported in 3HSP, e.g. 5BZG\,J1103+0022, 5BZB\,J1254+2211, 5BZG\,J2211-0023 and 5BZG\,J2248-0036. Considering that there could be many more sources with currently available X-ray data, we cross-match the ML-Blazar sample with several X-ray catalogues. The catalogues considered are (in priority order) 3XMM DR8, Swift-1SWXRT, Swift-XRTGRB , XMM-SL2, 2RXS-RASS, and Chandra V1.1. \citep{3XMM-DR8,SWIFT-XRT1,SWIFT-deep,XMMSL1,2RXS-RASS,Chandra}. The cross-match radius is defined in Table \ref{tab:xray-rad}, considering the position-error ($\rm p_{err}$) associated to each X-ray source. We group the X-ray sources according to its error radius in each catalogue and define a fixed radius for the cross-match between ML-Blazars and X-ray sources. 

\begin{table}
	\centering
	\caption{Cross-match radius for each X-ray catalogue, adapted according to the position-error $\rm p_{err}$ reported for each source. Columns describe the catalogue name, the total number (N) of sources in each of them, the position error range $\rm p_{err}$ (used to group sources before cross-matching with the ML-Blazar sample), and the fixed radius assumed for the cross-match.}
	\label{tab:xray-rad}
	\begin{tabular}{lccr} 
		\hline
		Catalogue & N & $\rm p_{err}$ & radius \\
		\hline
		2RXS-RASS     & 135.118 & all-data & 40'' \\              
        Swift-1SWXRT  & 84.979  & 0''-5''  & 0.1' \\
                      &         & $>$5''   & 0.2' \\
        Swift-XRTGRB  & 151.524 & all-data & 0.2' \\ 
        3XMM-DR8(2018)& 531.454 & 0''-5''  & 0.1' \\
                      &         & $>$5''   & 0.2' \\ 
        XMMSL2(2017)  & 29.393  & all-data & 10'' \\ 
        Chandra-V1.1  & 106.586 & all-data & 0.1' \\ 
		\hline
	\end{tabular}
\end{table}

This follows the same strategy as of the 1\&2WHSP and 3HSP blazar catalogues \citep{Arsioli2015,Chang2017,Chang2019} and helps to avoid losing relevant X-ray information. A total of 362 sources gained X-ray flux. Still, 1122 sources in the ML-Blazar sample had no measured X-ray flux ($\sim$25\%). Considering that RASS is the only all-sky survey available, and given that it covers the majority of the sky out of the galactic plane with flux-limit ranging from 10-1$\times$10$^{-13}$ erg/cm$^2$/s at 1 keV (flux density of $\sim$4-0.4$\times$10$^{-30}$ ergs/cm$^2$/s/Hz), we add an X-ray upper limit of 5.0$\times$10$^{-13}$ erg/cm$^2$/s ($\sim$2.0$\times$10$^{-30}$ erg/cm$^2$/s/Hz) to all cases with no X-ray counterpart. To keep easy track of the UL value, we add the X-rayUL attribute with value 1 for UL, and 0 for measured flux.

\subsection{Feature selection and Standardisation}
\label{sec:feature-selection}

\begin{table}
	\centering
	\caption{List of mean values and standard deviations associated with each feature, as used for its standardisation (eq. \ref{eq:std}). From the centre to the right columns, the KS test statistic D and the t-test statistics, with their corresponding p-values, to compare the feature's distributions for HSP and LSP sources. The best performing features for the classification task are marked with *.}
	\label{tab:std}
	\begin{tabular}{l|cc|cc|cc} 
	Features         & Mean   & $\sigma$   & KS D  &  p-value & t-test & p-value  \\ 
	\hline
	\textbf{Flux Density}  &  &          &        &         &        &         \\
	log(F Radio)  & -23.98   & 0.735     & 0.66   &  0.0    &   56.1 & 0.0     \\   
    log(F IR-w1)  & -26.15   & 0.530     & 0.11   &  2E-10  &   4.7  & 2E-6    \\ 
    log(F IR-w2)  & -26.07   & 0.529     & 0.13   &  7E-15  &   5.9  & 4E-9    \\ 
    log(F IR-w3)  & -25.82   & 0.530     & 0.51   &  2E-205 &   27.8 & 4E-152  \\  
    log(F Opt)    & -26.63   & 0.563     & 0.17   &  2E-24  &   10.9 & 5E-27   \\     
    log(F X-ray)  & -29.67   & 0.466     & 0.46   &  7E-161 &   25.0 & 4E-120  \\ 
    \hline
    \textbf{Radio-IR}  &      &           &        &           &        &          \\
    alpha-rw1*    &  0.448    & 0.170     & 0.71   &  0.0      &  62.3  &  0.0     \\
    alpha-rw2     &  0.445    & 0.159     & 0.66   &  0.0      &  54.7  &  0.0     \\
    alpha-rw3     &  0.429    & 0.145     & 0.51   &  3E-205   &  37.7  &  3E-263  \\
    \hline
    \textbf{Radio-O\&X}  &     &           &        &           &       &            \\
    alpha-ro*     &  0.484     & 0.162     & 0.71   &  0.0      & 64.1  &  0.0       \\
    alpha-rx*     &  0.686     & 0.111     & 0.76   &  0.0      & 66.7  &  0.0       \\ 
    \hline
    \textbf{IR-colours}   &     &           &        &           &       &            \\
    alpha-w2w1*   &  0.572     & 1.046     & 0.72   &  0.0      &  53.8 &  0.0       \\   
    alpha-w3w2    &  0.609     & 0.621     & 0.71   &  0.0      &  48.6 &  0.0       \\ 
    alpha-w1w3*   &  0.600     & 0.664     & 0.77   &  0.0      &  58.1 &  0.0       \\
    \hline
    \textbf{IR-Optical}   &    &           &        &           &       &            \\
    alpha-w1o     &  0.751     & 0.595     & 0.27   &  5E-58    &  9.4  &  1E-20     \\
    alpha-w2o     &  0.722     & 0.527     & 0.43   &  4E-147   &  25.2 &  1E-129    \\
    alpha-w3o     &  0.683     & 0.433     & 0.66   &  0.0      &  48.1 &  0.0       \\
    \hline
    \textbf{IR-Xray}  &        &           &        &           &       &           \\
    alpha-w1x     &  0.751     & 0.595     & 0.24   &  2E-44    &  17.2 & 1E-62     \\ 
    alpha-w2x     &  1.002     & 0.166     & 0.37   &  8E-108   &  26.4 & 7E-135    \\ 
    alpha-w3x     &  0.961     & 0.164     & 0.56   &  2E-245   &  44.6 & 3E-315    \\ 
    \hline
    \textbf{Opt-Xray}  &   &               &        &           &       &           \\
    alpha-ox      &  1.076     & 0.204     & 0.16   &  5E-20    &  11.5 & 4E-30     \\   
	\hline
	\end{tabular}
\end{table}

Following the preprocessing stage, the ML-Blazar data-frame has six flux attributes ($f$: radio, IR-w1, IR-w2, IR-w3, optical, X-ray), and two main categorical flags (source-type and X-rayUL) associated to each source, summing a total of 8 attributes. To try and condensate multifrequency information into fewer parameters, we calculate spectral slopes $\alpha_{\nu_1 \nu_2}$ between bands according to: 

\begin{equation}
   \alpha_{\nu_1 \nu_2} \text{=} - \frac{\log \left( f_{\nu_1} \text{/} f_{\nu_2} \right) }{\log \left( \nu_1 \text{/} \nu_2 \right) }
\label{eq:slope}   
\end{equation}

The spectrum slopes are well representative of the energy distribution (SED shape) needed for our classification, and carry information from two bands into a single $\alpha$ parameter. We test the possibility to reduce the number of attributes necessary to train our ML model, by working with slopes instead of fluxes. However, at this stage, any combination of bands (into $\alpha$) could turn out relevant, so that we add another fifteen attributes to the data-frame for evaluation purpose: $\rm \alpha_{rw_1}$, $\rm \alpha_{rw_2}$, $\rm \alpha_{rw_3}$, $\rm \alpha_{ro}$, $\rm \alpha_{w_1 w_2}$, $\rm \alpha_{w_1 w_3}$, $\rm \alpha_{w_2 w_3}$, $\rm \alpha_{w_1 o}$, $\rm \alpha_{w_2 o}$, $\rm \alpha_{w_3 o}$, $\rm \alpha_{rx}$, $\rm \alpha_{w_1 x}$, $\rm \alpha_{w_2 x}$, $\rm \alpha_{w_3 x}$, and $\rm \alpha_{ox}$. By defining slopes between channels, we wish to retain information within a small number of attributes. We look forward to reducing the dimension of the problem by keeping only the ones that are more promising to separate between blazar classes.

We perform the standardisation (std) of all features, to represent them as normal distributions with 0.0 mean value and unitary standard deviation ($\rm \sigma$). To do that, we take each input value, e.g. $\alpha_{\nu1-\nu2}$ , subtract it by the feature-mean $\rm \langle \alpha_{\nu1-\nu2} \rangle$, and divide by the feature standard deviation $\rm \sigma_{\alpha_{\nu1-\nu2}}$, eq. \ref{eq:std}. The mean and $\rm \sigma$ values are calculated for the feature distribution considering all sources together (meaning: there is no need to separate HSP and LSP classes) and are listed in Table \ref{tab:std}. Standardisation is important for the optimisation of the machine learning algorithms because models can misleadingly assign a large weight (importance) to a given feature, only because of its dominant absolute value. Therefore, the standardisation is commonly applied to level-out all features, so that all have the same weight when computing cost functions during the optimisation process \footnote{Standardisation is especially relevant for algorithms which rely on the calculation of Euclidean distances between points in multidimensional space (Support Vector Machine, K-neighbours), or when gradient descent is used for optimisation in Neural Networks.}.    

\begin{equation}
\log \left( f \right)_{std}  \text{=}  \frac{\log \left( f \right) - \langle \log \left( f \right) \rangle}{\sigma_{\log \left( f \right) }} \   ,  \   \alpha_{std}  \text{=}  \frac{ \alpha_{\nu_1 \nu_2} - \langle \alpha_{\nu_1 \nu_2} \rangle}{\sigma_{\alpha_{\nu_1 \nu_2} }} 
\label{eq:std}
\end{equation}

A Kolmogorov-Smirnov test \citep[KS,][]{KS-Test-Hodges1958} is applied to evaluate if the HSP and LSP samples are drawn from the same distribution (the null hypothesis H$_0$). The KS test statistic D is the maximum absolute difference between the two cumulative distribution functions, and the corresponding two-sided p-value is the probability that we would see a $\geq$D value, only by chance, given that H$_0$ is true. A two-sided p-value\,<\,0.05 is usually interpreted as a denial of H$_0$, meaning that the LSP and HSP distributions are not drawn from the same population. Using the scipy.stats.ks-2samp library \citep{Scipy}, the p-values obtained for all spectral sloes (Table \ref{tab:std}) reject H$_0$ at <1\% level, meaning that the distributions are not drawn from the same populations. In other words, all features $\rm \alpha_{\nu_1 \nu_2}$ hold unique information about the LSP and HSP classes. A collection of five slopes -$\rm \alpha_{rw_1}$, $\rm \alpha_{ro}$, $\rm \alpha_{rx}$, $\rm \alpha_{w_2 w_1}$, $\rm \alpha_{w_1 w_3}$- enclose information along the radio to the X-ray band, with the largest KS D values (all >0.7) and the lowest P$_{values}$ (all $\sim$0.0). The largest D values associated to nearly null p-values are a strong indication that those five slopes are the most relevant parameters to separate/classify blazars into LSP or HSP type. 

\begin{figure}
    \centering
    \includegraphics[width=1.0\linewidth]{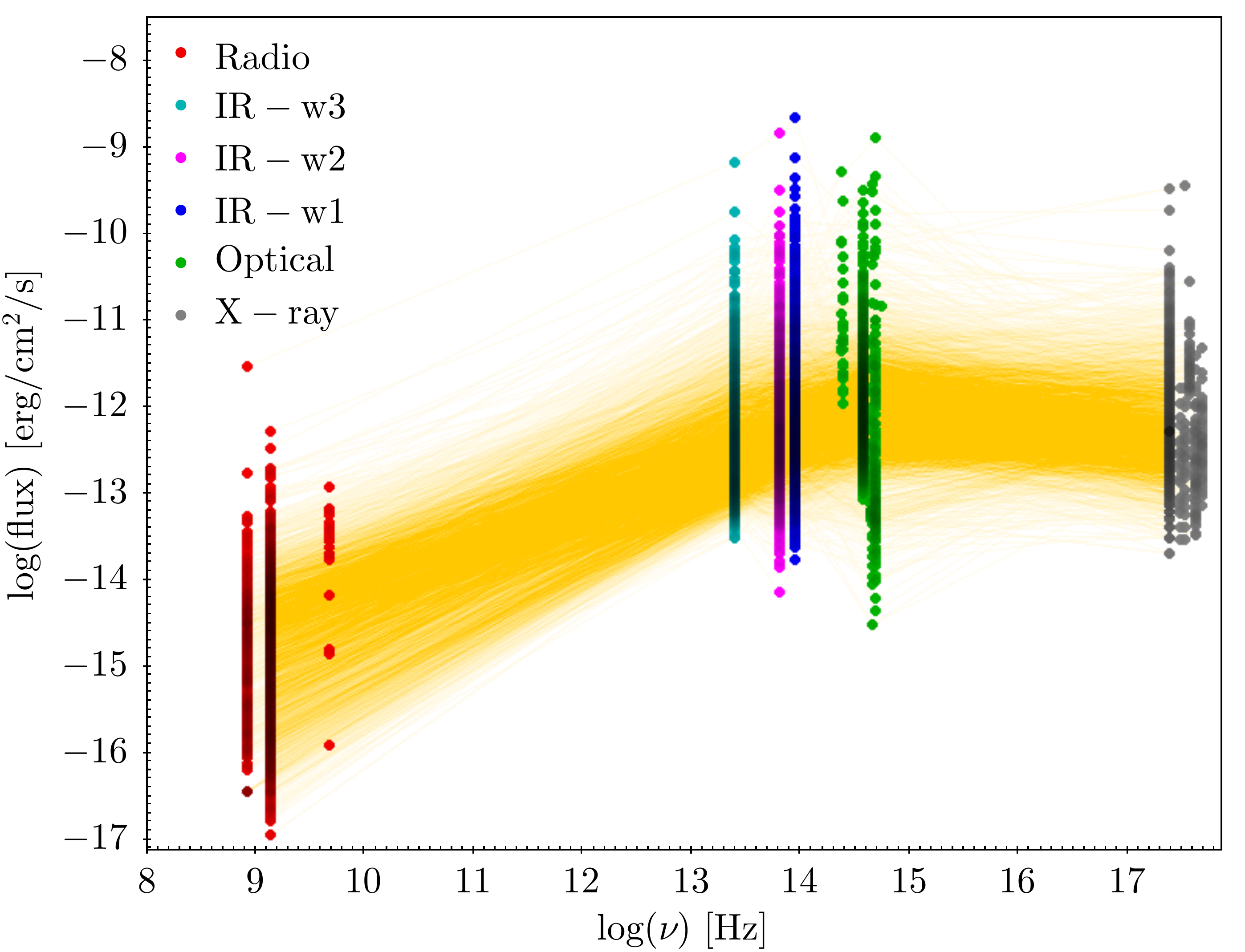}
    \caption{A superposed SED of all sources of the ML-Blazar data-frame. From left to right, the bands are: Radio (red), IR w1(cyan) w2(magenta) w3(blue), Optical (green), and X-ray (grey). The fluxes corresponding to each blazar were connected with yellow translucent lines to form a density view of the parameter space covered by the entire data-frame.}
    \label{fig:entire-sed}
\end{figure}

\begin{figure*}
    \centering{
    \includegraphics[width=1.0\linewidth]{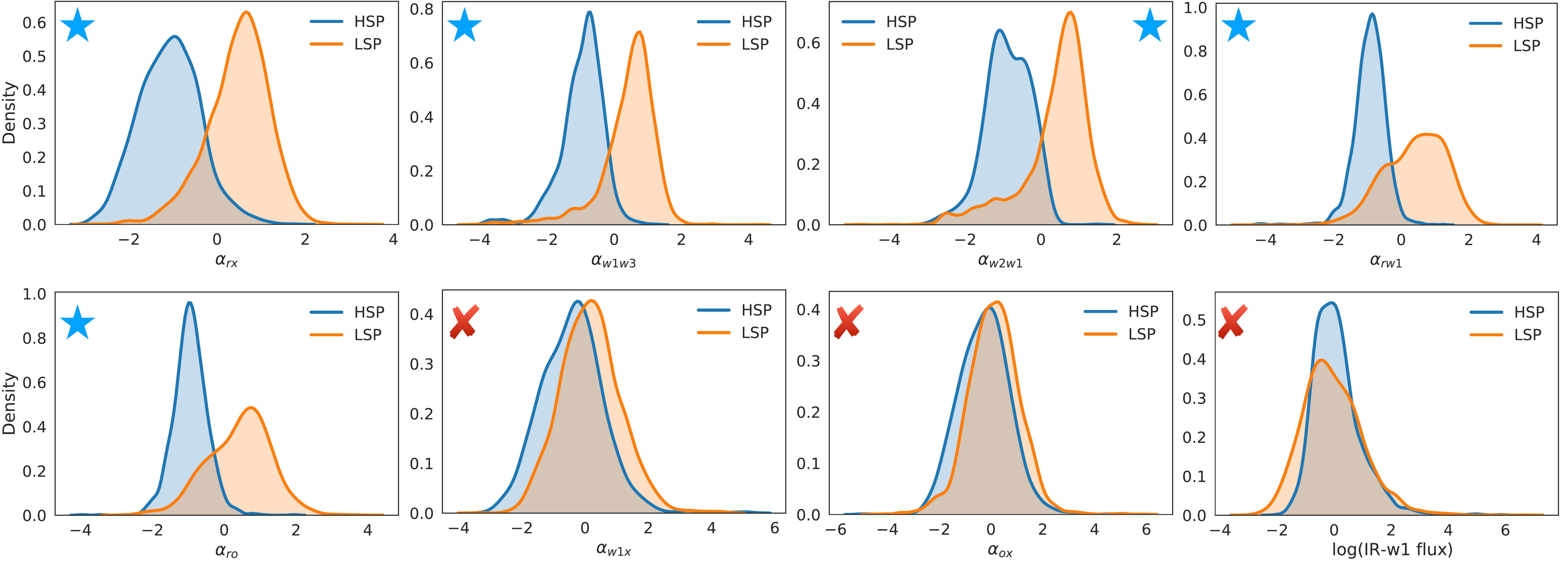}
    \caption{Density plots with the distribution of standardised values per feature, and dividing between HSP (blue) and LSP (orange) sources. The five spectral slopes selected for the ML training (with the largest KS D, and lower p-values) $\rm \alpha_{rw_1}$, $\rm \alpha_{ro}$, $\rm \alpha_{rx}$, $\rm \alpha_{w_2 w_1}$, and $\rm \alpha_{w_1 w_3}$ are marked with blue stars, and another three examples of features with the lowest KS D values, marked with a red cross.}}
\label{fig:matrix-alpha}
\end{figure*}

We also apply a t-test for independent-samples to compare the mean values of a given feature and for each class. It measures how separable are the classes according to the means, and the absolute t-test value can be used to compare the features (given that all are standardised). According to the t-test, all features are distinct with respect to the mean, at high confidence level, p-values<1\%; and the six best performing features (most separable according to the mean) are: $\rm \alpha_{rx}$, $\rm \alpha_{ro}$, $\rm \alpha_{rw_1}$, $\rm \alpha_{w_1 w_3}$, $\rm \alpha_{r w_2}$, and $\rm \alpha_{w_2 w_1}$. Those enclose all the five slopes highlighted by the Kolmogorov-Smirnov test, therefore pointing to a similar conclusion.

From now on, we focus on these five spectral slopes ($\alpha$: rw1, ro, rx, w2w1, and w3w2) to train and optimise several machine learning algorithms while keeping a low number of features. The dimension of the problem reduces from seven input features (six flux density + one X-ray upper limit flag) to only five spectral slopes. Moreover, the statistical tests show that spectral slopes can separate blazar classes better than the fluxes alone (as seen by the results in table 3). 

Fig.~\ref{fig:matrix-alpha} shows the density distribution of HSPs and LSPs for the five selected alphas, as a primary assessment of how separable are LSP and HSP sources based on each feature. This rather qualitative view is complementary to the results reported in Table \ref{tab:std}, which lists the Kolmogorov-Smirnov test statistics D and the t-test statistics, applied to compared the LSP and HSP distributions. 


\section{The Machine Learning Algorithms and Metrics}
\label{sec:MLalgos}

Here we briefly describe the machine learning algorithms used in this work (Random Forest (RF), Support Vector Machine (SVM), K-Nearest Neighbours (KNN), Gaussian Naive Bayes (GNB), and Ludwig) and the configuration parameters which are available to optimise each of them. Following, we discuss the metrics used to evaluate the performance of the ML models, and the re-sampling strategy applied to validate and access the uncertainty associated with each of them.   

{\bf Random Forest (RF):} The Random Forest\footnote{\href{https://scikit-learn.org/stable/modules/generated/sklearn.ensemble.RandomForestClassifier.html?highlight=random\%20forest\#sklearn.ensemble.RandomForestClassifier}{Scikit-learn RF documentation.}} is an extension of the Decision Tree method. Decision Trees are supervised machine learning algorithms used for classification based on decision rules. These decisions are organised hierarchically in nodes, where a decision is made based on the feature that best separates the classes in that node using a given metric (either gini indices or entropy). The tree starts on the so-called root node, made by the feature with the best dividing metric for the whole data, and ends on leaves where there are no better divisions to be made. To work around the problem of biased Decision Trees, the Random Forest uses an ensemble of trees, each one built from a random sample of the training set (bootstrap sample). Also, instead of using the best split among all the features, the choice is made among a random subset of features. Therefore, in this kind of algorithm, the most critical optimisation parameters are the number of trees (or estimators), the size of the features subset and the tree's maximum depth. Finally, each tree of the Random Forest counts as a vote for the classification of each object, and the ensemble of votes defines the final classification.  

{\bf Support Vector Machine (SVM):} The Support Vector Machine classifier\footnote{\href{https://scikit-learn.org/stable/modules/generated/sklearn.svm.SVC.html\#sklearn.svm.SVC}{Scikit-learn SVM documentation.}} uses hyperplanes on the features space trying to best separate the objects among their classes. The algorithm maximises the distance between the hyperplane and the objects of any class, trying to minimise classification errors. One attribute of this kind of algorithm is the Kernel, that is a transformation from the features space to another space that better separate the classes. The most used Kernels are the linear, polynomial (poly), radial basis function (RBF) and sigmoid. For all the SVM kernels there is a regularisation parameter called C, which is a penalty for the wrong classification. A high C forces the model into higher precision, but it could also bias towards overfitting the training data. For the poly, RBF and sigmoid, there is the gamma parameter, which is a scale factor for the computed distances between the hyperplane -in the feature space- and the objects of a given class. And for the poly kernel, variations of the polynomial degree can be explored to improve modelling.

{\bf K-Nearest Neighbours (KNN):} The K-Nearest Neighbours algorithm\footnote{\href{https://scikit-learn.org/stable/modules/generated/sklearn.neighbors.KNeighborsClassifier.html?highlight=kneighbors\#sklearn.neighbors.KNeighborsClassifier}{Scikit-learn KNN documentation.}} implements a voting system where the class of the nearest neighbours objects define the classification. In the simplest implementation, the algorithm computes the object class by the majority of votes where all the k nearest neighbours have the same weight. Another approach is to weight the neighbour vote by the inverse of its distance from the object. In both cases, by distance, we mean the Euclidean distance, which can be computed in multidimensional parameter space. Therefore, the main variables to play in this model are the number of neighbours used in the classification and the weight for them, uniform or radius dependent.

{\bf Gaussian Naive Bayes (GNB):}
The Naive Bayes\footnote{\href{https://scikit-learn.org/stable/modules/generated/sklearn.naive_bayes.GaussianNB.html\#sklearn.naive_bayes.GaussianNB}{Scikit-learn GNB documentation.}} classifiers use the Bayes Theorem from statistics to predict the class given the features of an object, making the "naive" assumption of conditional independence among the features to simplify the calculations. The result of this assumption is a simple formula:

\begin{equation}
P\left( y|x_{1} ... x_{n} \right) \propto P\left( y  \right) \prod_{i=1}^{n} P\left(x_{i}|y\right),
\end{equation}
where $y$ is the class and $x_{i}$ the ith feature of a given object. The Gaussian Naive Bayes algorithm is the one that assumes $P(x_{i}|y)$ to follow a Gaussian distribution. Also, $P\left( y  \right)$, called prior probability, is the probability of each class in the dataset. To predict the class, the GNB algorithm chose $y$ that maximise the probability that the object belongs to that class. Although this type of algorithm uses simplified assumptions, it produces good results in a variety of applications, besides being one of the lightest algorithms to run.

{\bf Ludwig:} Ludwig is an open-access framework of deep learning, able to deal with  classification-problems, and designed to be autonomous in fine-tuning its internal parameters to accomplish efficient and accurate classification of objects \citep{ludwig-2019}. Ludwig is made available by the Uber team\footnote{\href{https://github.com/uber/ludwig}{The ludwig:github documentation}.}, and we use it to compare its automated modelling with our optimised results. This way, we access the power of Ludwig for science use cases, noting it is versatile and relatively easy to implement.

\subsection{Statistical Analysis \& Evaluation Metrics}

One of the most common metrics used to evaluate the model response is the accuracy \cite[\textit{accuracy\_score},][]{scikit-learn}, defined as the fraction of objects correctly classified by the model (the number of true positives TP, divided by the total number of objects in the sample, n$_{samples}$) which gives us a performance response independent of class. To complement the analysis, two other metrics -recall and precision- are commonly considered for model evaluation; In a `per class' perspective, those represent the completeness of a class classified by the ML model and the efficiency of the classifier. The precision is defined as the number of true positives (TP) divided by the number of elements classified as belonging to the class (True Positives + False Positives: TP+FP); the recall is defined as the true positives (TP) divided by the number of elements in the class (True Positives + False Negatives: TP+FN) \citep[\textit{classification\_report},][]{scikit-learn}). The precision tells how clean is the classification for each given class, and the recall tells the fraction of objects of a class that were classified correctly. 

\begin{figure}
\centering
\includegraphics[width=1.0\linewidth]{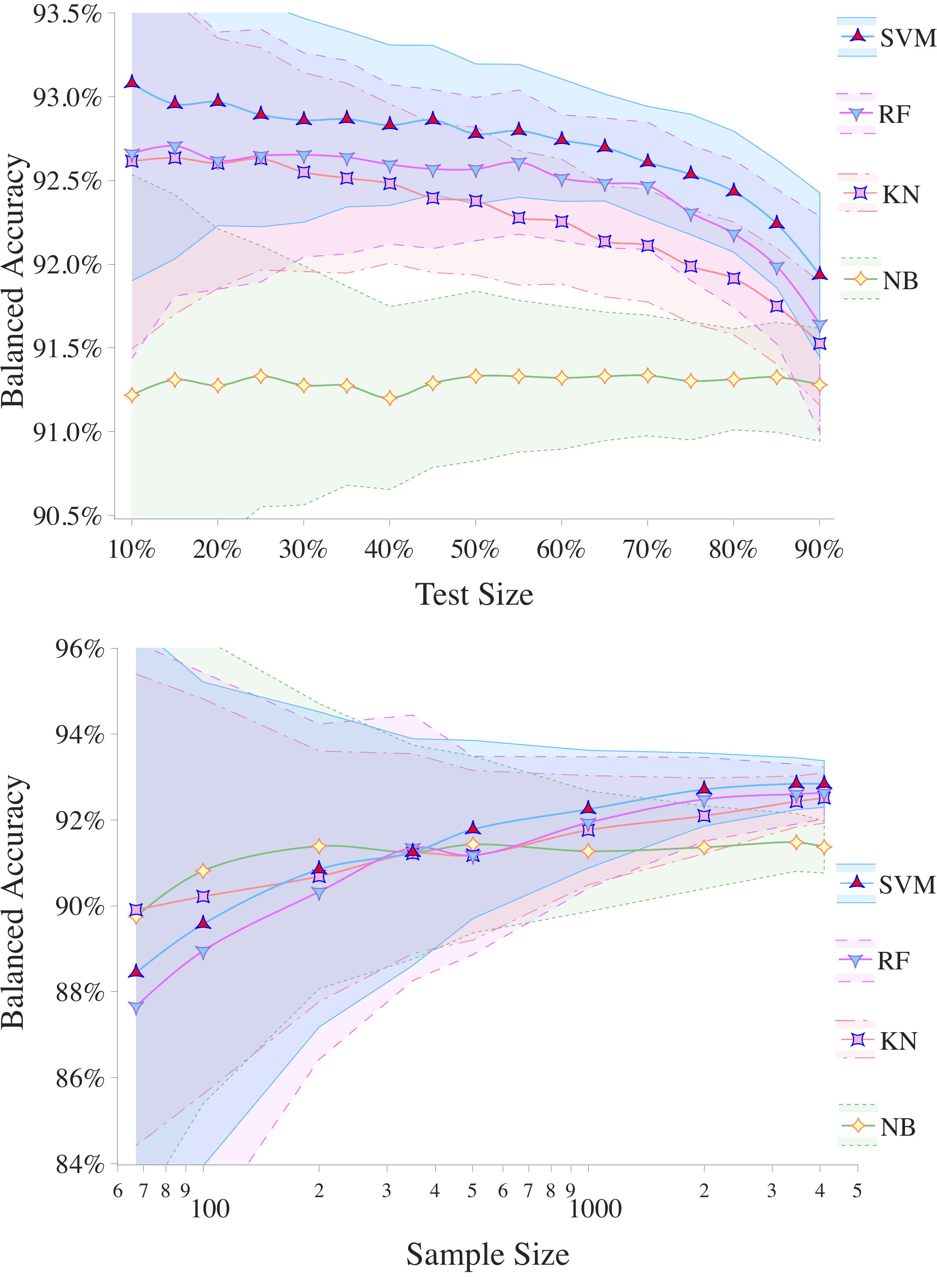}
\caption{Top: Balanced-accuracy dependence on the test sample size given in fractions of the total dataset size. Bottom: Balanced-accuracy dependence with respect to the sample-size. This results correspond to the algorithms: RF, SVM, KNN, and GNB; all trained with the optimal fitting parameters for each algorithm (as described in Sec. \ref{sec:tunning}).}
\label{fig:TestSampleSize-vs-Accuracy}
\end{figure}

Another metric commonly used to evaluate the performance of categorical classifiers is the f1-score, which represents the weighted average between precision and recall for each class. The use of f1-score has its limitations since it gives equal weight to precision and recall, but it can be used to compare models in terms of the optimum balance between precision and recall. 

However, for an imbalanced data set, both the accuracy and f1-score can not provide the full picture, because the metric gets biased towards the dominant class, i.e. the LSPs. To account for the imbalance we look for the scikit-learn implementation of the  \citep[\textit{ balanced\_accuracy\_score},][]{Brodersen-BalAcc-2010} that correspond to the average of recall for each class; and the weighted-average f1-score\footnote{About the balanced (weighted) f1-score, check \href{https://scikit-learn.org/stable/modules/generated/sklearn.metrics.f1_score.html}{the f1 metrics documentation on scikit-learn}.} that calculates both the precision and recall metrics for each label, and weight the f1-score based on the number of objects in each class. All metrics listed were used to evaluate the models regarding the test samples: 

\begin{equation}
\rm Accuracy = \frac{TP + TN}{TP + FP + TN + FN} = \frac{TP_{all-classes}}{n_{samples}}   
\end{equation}

\begin{equation}
\rm Bal-Accuracy = \frac{1}{2} \Big( \frac{TP_{class-A}}{n_{class-A}} + \frac{TP_{class-B}}{n_{class-B}} \Big) 
\end{equation}

\begin{equation}
\rm Precision = \frac{TP}{TP + FP}; \ per \ class
\end{equation}

\begin{equation}
\rm Recall = \frac{TP}{TP + FN} ; \ per \ class
\end{equation}

\begin{equation}
\rm f1 \ score = \frac{2\times Precision \times Recall }{Precision+Recall}; \ per \ class
\end{equation}

\begin{equation}
\rm Weighted \ f1 = \frac{n_{class-A} \times f1_{class-A} + n_{class-B} \times f1_{class-B}}{n_{samples}} 
\end{equation}

In the perspective of population studies in astrophysics, precision and recall translates respectively into selection-efficiency and sample-completeness. Both completeness and efficiency are highly valuable for astrophysics, but the trade-off between them may depend on the application \citep{DataMine-ML-Astronomy-2010}. In case of searching for rare objects as blazars within large multifrequency databases, the most effective strategy would likely be to give priority to completeness (recall). Allowing lower efficiency -which leads to a certain degree of contamination- can be remedy latter by additional strategies, as cleaning the final database with hand-implemented cuts, or even by human intervention via case-by-case investigation of border-class cases. 

\begin{figure}
    \centering
    \includegraphics[width=1.0\linewidth]{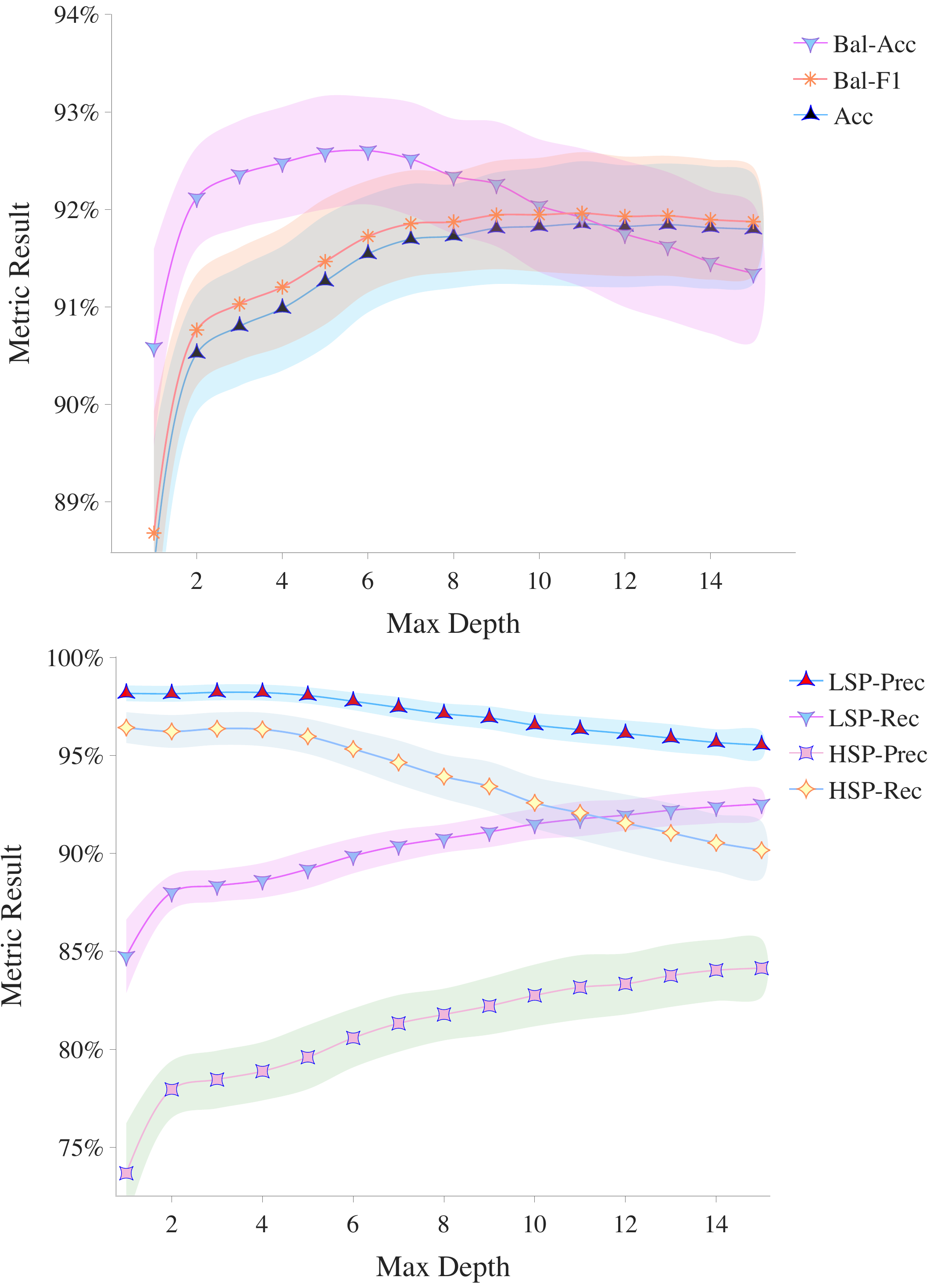}
    \caption{RF Metrics. Top: The balanced-accuracy, balanced-f1, and accuracy metrics with n-features=1 and n-estimators=250, setting the weight option to balance the training sample. Bottom: The precision and recall of the HSP and LSP classes. All metrics are shown as a function of the max-depth parameter, that ranges from 1 to 15. The points and uncertainty bands are the mean and $\sigma$ for each metric, considering 300 re-samples.}
    \label{fig:RF_Parameter_Tuning_Oversampled}
\end{figure}

\begin{figure}
    \centering
    \includegraphics[width=1.0\linewidth]{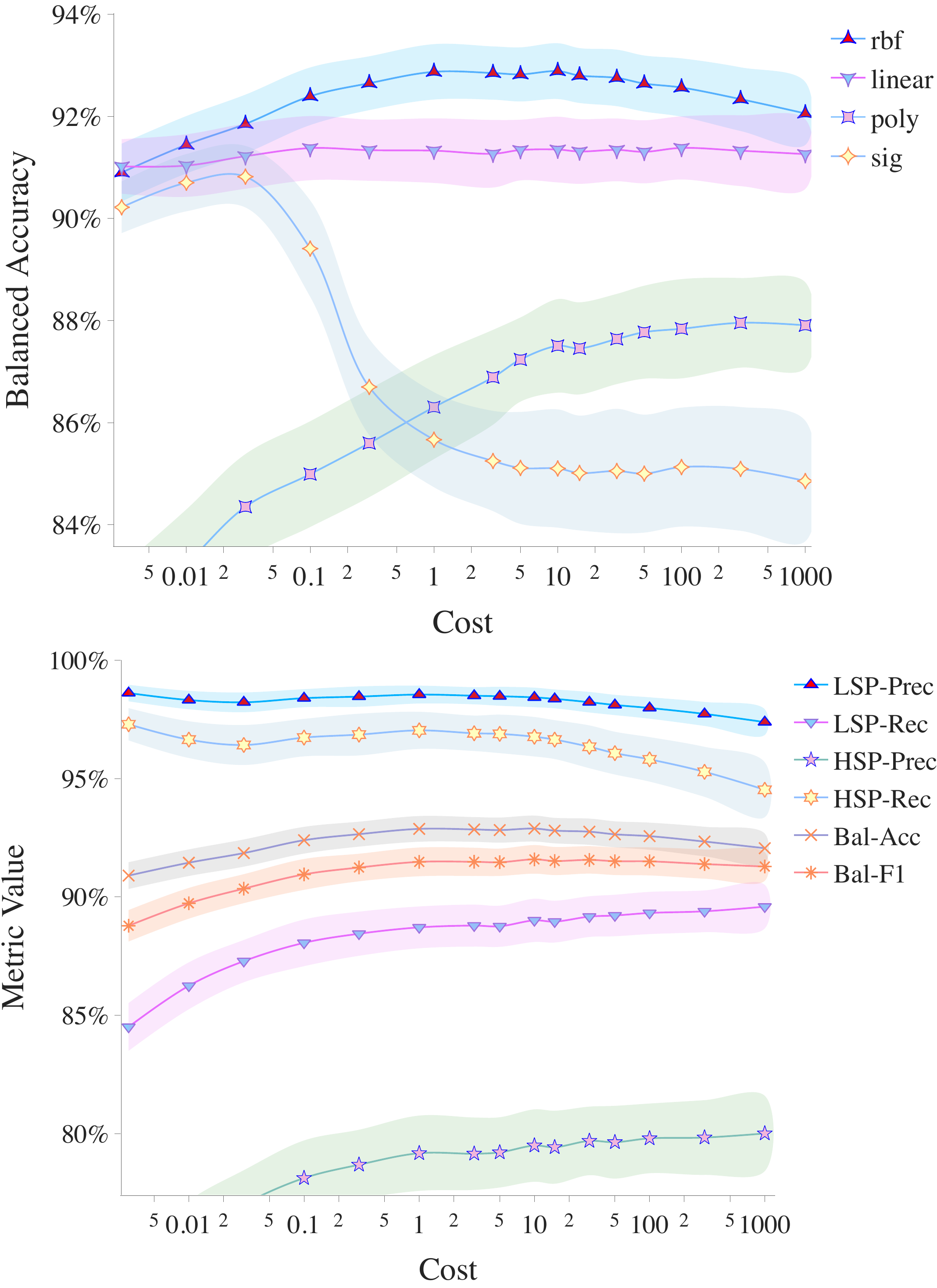}
    \caption{SVM Metrics. Top: The balanced-accuracy for the rbf, linear, polynomial (poly), and sigmoidal (sig) kernels. Bottom: Compilation of the SVM metrics considering the rbf kernel, showing precision and recall for the HSP and LSP classes, together with the balanced-f1 and balanced-accuracy scores. All metrics are shown as a function of the Cost parameter (C), that ranges from 0.003 to 1000. The points and uncertainty bands are the mean and $\sigma$ for each metric, considering 300 re-samples.}
    \label{fig:SVM-Kernel-and-C-dependency}
\end{figure}

\subsection{Re-sampling \& Validation}

To analyse the results of the classifier, a commonly used tool in machine learning is the so-called K-fold cross-validation, where the data set is divided into K parts, with one of these used as the testing set and the rest used for training. The process is repeated for each K part of the testing set to calculate the average over the performances. Our approach is similar, but instead of pre-dividing the data set, we split it for training and testing using a pseudo-random algorithm and repeat this process 300 times, so that each time we have a randomly selected train and test sets. The final result is a distribution of the performance, where we can extract the mean and standard deviation ($\sigma$) associated with any metric. The degree of modelling uncertainty will directly translate into large/small $\sigma$. A relatively small $\sigma$ signals that the trained models are stable regarding randomly selected train and test samples, and up to the classification task for real applications.

\begin{figure}
    \centering
    \includegraphics[width=1.0\linewidth]{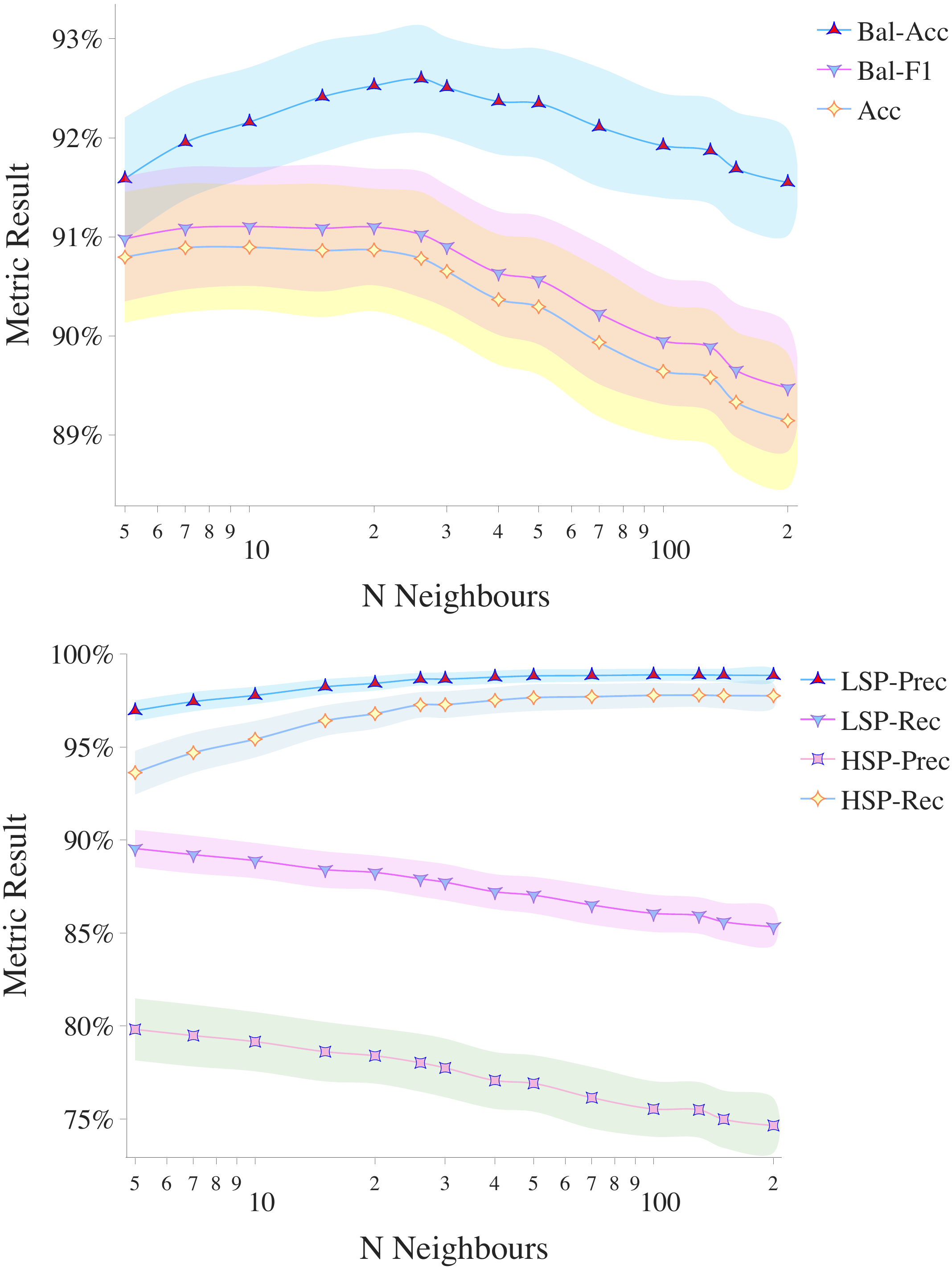}
    \caption{KNN Metrics. Top: The balanced-accuracy, balanced-f1, and accuracy metrics as a function of the number of neighbours; using the distance weights kernel, and considering the SMOTE algorithm to balance the training sample. Bottom: The precision and recall of the HSP and LSP classes. The points and uncertainty bands are the mean and standard deviation for each metric, considering 300 re-samples.}
    \label{fig:KN_N_Neighbors}
\end{figure}

\begin{figure}
    \centering
    \includegraphics[width=1.0\linewidth]{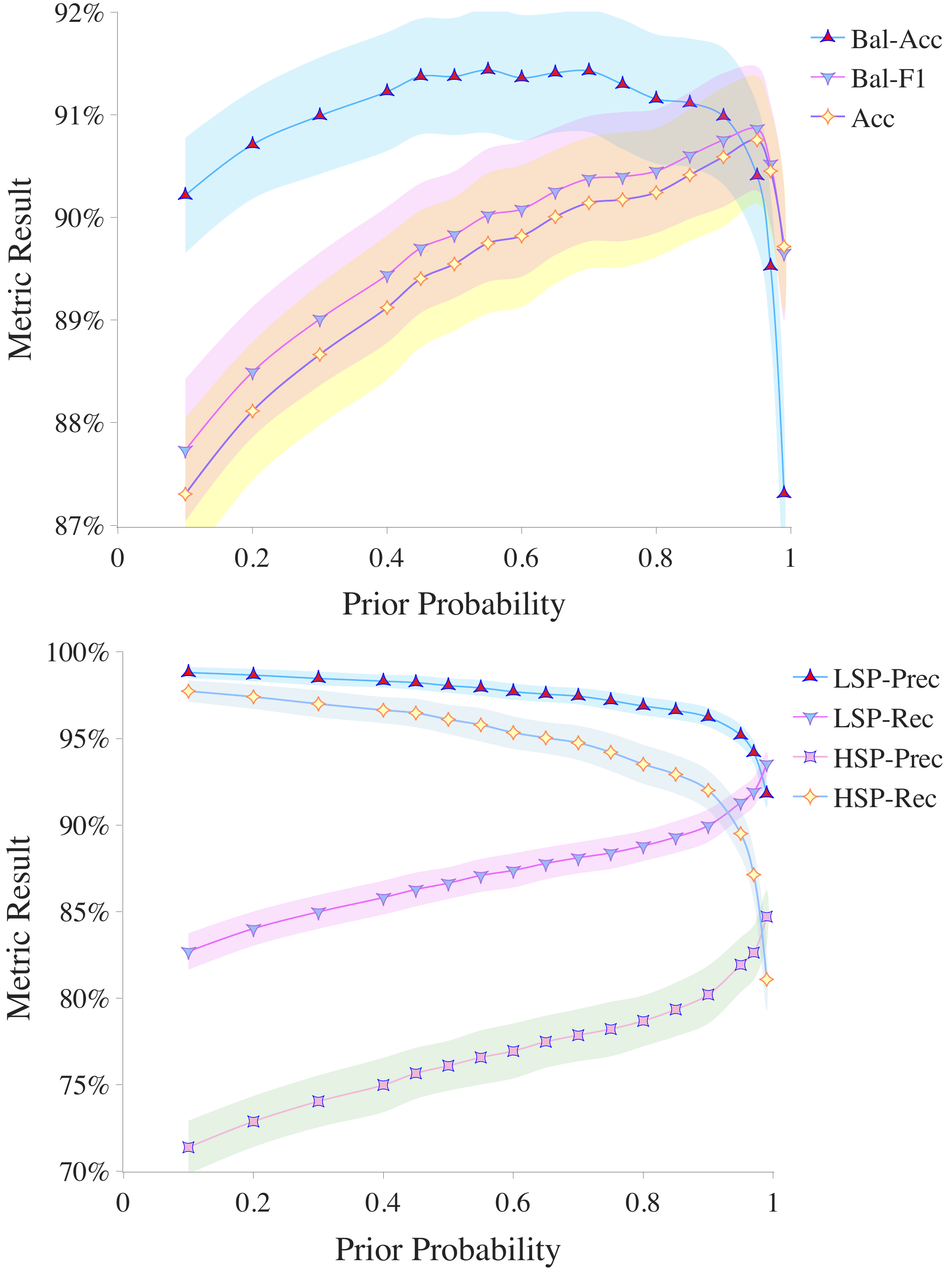}
    \caption{GNB Metrics. Top: The balanced-accuracy, balanced-f1, and accuracy metrics as a function of the Prior Probability (for the LSP class), and considering the SMOTE algorithm to balance the training sample. Bottom: The precision and recall of the HSP and LSP classes. The points and uncertainty bands are the mean and standard deviation for each metric, considering 300 re-samples.}
    \label{fig:GNB_Prior_Prob}
\end{figure}

\section{Discussion and Results}
\label{sec:MLoptmization}

Here we walk through the steps of model optimisation for the RF, SVM, KNN, and GNB algorithms. Each ML algorithm has a different space of fitting parameters to probe, and we scan over them to select the best configuration focusing on the balanced-accuracy metric. For the Ludwig framework, we only feed the data and collect results, with no optimisation other than its auto-ML functionality.

\begin{figure*}
    \centering
    \includegraphics[width=1.0\linewidth]{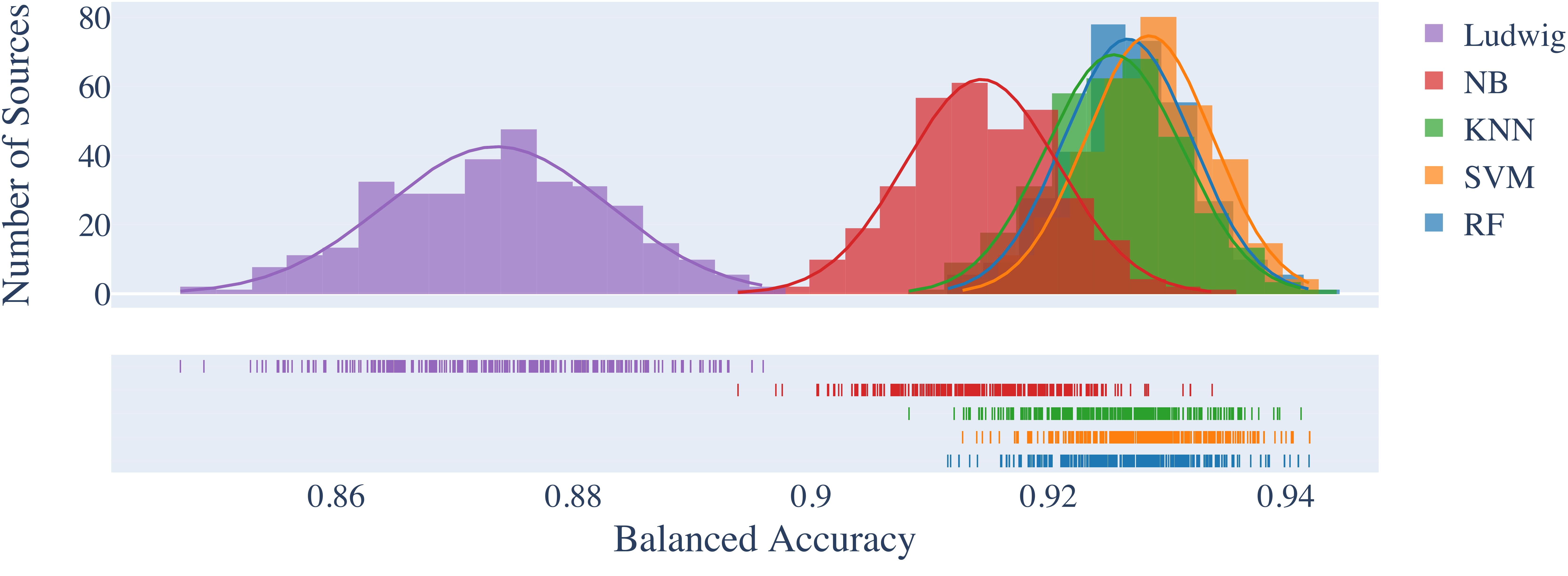}
    \caption{Balanced-Accuracy distribution for all the ML algorithms with optimal configuration, including results for the Ludwig auto-ML tool. The bottom density-box shows the distribution of individual bal-acc values.}
    \label{fig:bal-acc-histo}
\end{figure*}

\subsection{optimising the train and test sample size}
\label{sec:train-test}

Before any attempt to optimise the ML models, we tried to understand which is the best splitting between train and test samples, and set the test size to vary between 10\% to 90\% of the total sample size. Initially, the ML algorithms run over default configuration \citep[as specified for each algorithm at sci-kit learn library,][]{scikit-learn} and using the balanced-accuracy as performance criteria. Later on, after we optimise the ML algorithms, the evaluation was performed again, and from that follows the results we report here\footnote{This results in a recurrence problem, where at each iteration the configuration converges to the best one. However, for this specific test, we obtained just minor corrections in the first iteration, and the selected train:test ratio remains the same, 0.65:0.35.}. Fig.~\ref{fig:TestSampleSize-vs-Accuracy} (top) shows the mean balanced-accuracy (bal-acc) and its standard deviation as a function of the test sample size, for the RF, SVM, KNN, and GNB models. The mean bal-acc and $\sigma$ are calculated from 300 re-samples. From there we decide to split the ML-Blazar sample into 65\% for training and 35\% for testing, to keep the test uncertainty at $\sim$1\% level for all models while keeping the mean balanced-accuracy as high as possible, and depleting the training sample as less as possible.

We also tested if our data-frame is large enough concerning the total number of objects (sample size). Using a random split, we reduced the dataset to smaller samples and trained the ML algorithms. At first, the algorithms were train \& test using the default configuration, and later the results were updated with the optimised versions (the ones described in Sec. \ref{sec:tunning}). Fig.~\ref{fig:TestSampleSize-vs-Accuracy} shows the bal-acc and $\sigma$ scanning the dataset sizes from 70 to 4000, and considering 300 re-samples for each point. As seen, the mean bal-acc evolves from 88-90\% up to 91-93\%, and the $\sigma$ reduces from a $\sim$15\% band down to $\sim$1\% band. Although the gains in bal-acc seem relatively small, the $\sigma$ shrinks drastically as the sample size grows. That means the uncertainty associated with each trained model is lower, and the the models are increasingly stable, and robust. By this analysis, we assure that our data-frame is large enough to produce models with small uncertainty and a high level of balanced-accuracy.

\subsection{Fine-tuning the ML Models}
\label{sec:tunning}

Here we describe the optimisation steps for each ML algorithms, scanning over the available parameters to fine-tune the models. We look forward to maximising the performance of the mean balanced-accuracy and report on the optimal setup conditions to meet that goal. We adopt the re-sampling as part of the uncertainty analysis and model validation \citep{Data-Intensive-Analysis-Kelling2009}, which showed to be a rather compute-intensive task. Our re-sampling analysis consists of 300 full train and test evaluations, running over randomly selected train \& test samples, to access the standard deviation of all target metric (e.g. the balanced-accuracy). 

The choice of the best performing model, however, depends on the starting goals. As an example, to maximise the LSP or HSP recall (completeness) without caring for the contamination of the sample, one could use different optimisation parameters. To illustrate those possibilities, in addition to the balanced-accuracy dependence, we show the precision, recall, and balanced-f1 scores. 

\textbf{Random Forest (RF):} The main parameters we optimise were: the number of features, the number of trees (or estimators), and the tree's maximum depth. We balance the training sample by assigning a larger weight to the less frequent class (the HSPs); setting the option \textit{class\_weight= `balanced'}. The number of features was scanned from 1 to 5, showing better performance at 1; the number of trees was scanned from 10 to 1000, performing best at 250; and the maximum depth was scanned from 1 to 15, performing best at 5, as shown in Fig.~\ref{fig:RF_Parameter_Tuning_Oversampled}. Those results emerge from an iterative procedure, starting with default values, and followed by fine-tuning, reaching an optimum balanced-accuracy of 0.926$\pm$0.006.

\textbf{Support Vector Machine (SVM):} We scan over the cost (C) value, for the range between 0.003 to 500, and each of the four kernels: rbf, linear, polynomial, and sigmoidal. We balance the training sample setting the option \textit{class\_weight= `balanced'}, which assign a larger weight to the the HSPs. The results are shown in Fig.~\ref{fig:SVM-Kernel-and-C-dependency}, with the rbf kernel reaching the best performance and optimum balanced-accuracy for C$\sim$8. We also test how the gamma parameter affects results, but the effect is negligible in our case; therefore, we set the `auto' option.

\begin{table*}
    \centering
    \caption{Mean performance metrics for each ML algorithm, for the setup that maximises the balanced-accuracy. The means and $\sigma$ are calculated based on 300 re-samples. The latest column to the right `Opt-Acc' corresponds to the optimum accuracy that could be reached with other setups.}
    \begin{tabular}{ |c|c|c|c|c|c|c|c|c|} 
     \hline
     Classifier & Bal-Acc & Bal-F1 & HSP-Recall &HSP-Precision &LSP-Recall &LSP-Precision& Acc & Opt-Acc   \\ 
     \hline
     RF   & $0.926\pm0.006$ & $0.915\pm0.006$ & $0.959\pm0.008$ & $0.796\pm0.015$ & $0.892\pm0.008$ & $0.980\pm0.007$ & $0.913\pm0.006$ & $0.919\pm0.006$  \\ 
     SVM  & $0.929\pm0.005$ & 0.915$\pm$0.006 & $0.968\pm0.008$ & $0.795\pm0.015$ & $0.890\pm0.009$ & $0.984\pm0.004$ & $0.914\pm0.006$ & $0.914\pm$0.006  \\ 
     KNN   & $0.926\pm0.006$ & $0.910\pm0.006$ & $0.973\pm0.007$ & $0.780\pm0.015$ & $0.879\pm0.009$ & $0.986\pm0.003$ & $0.908\pm0.006$ & $0.909\pm0.006$ \\ 
     GNB & $0.914\pm0.006$ & $0.903\pm0.006$ & $0.946\pm0.010$ & $0.777\pm0.015$ & $0.880\pm0.008$ & $0.974\pm0.005$ & $0.897\pm0.007$ & $0.907\pm0.006$ \\
     Ludwig & $0.873\pm0.015$ & $0.887\pm0.012$ & $0.837\pm0.029$ & $0.801\pm0.024$ & $0.908\pm0.012$ & $0.926\pm0.012$ & $0.886\pm0.011$ & $0.886\pm0.011$ \\
     \hline
    \end{tabular}
    \label{tab:model-metrics}
\end{table*}

\textbf{K-Neighbors(KN):} In KNN algorithm, the main optimisation parameter is the number of neighbours, which we scan from 5 to 200, as shown in Fig.~\ref{fig:KN_N_Neighbors}. We have tested both kernels available to weight the k neighbours, `distance' and `uniform', which shows slightly better results (~0.2\% more accurate, on average) for the `distance' kernel. Given the unbalance does affect the KNN modelling \citep[as reported in ][]{KNN-unbalanced-2012}, we implement a SMOTE strategy to balance the training sample (within each re-sample loop) delivering an overall improvement of $\sim$0.5\% in balanced-accuracy. Here we report results for the kernel weight `distance', and for the training-sample balanced with SMOTE. There is an optimum region for n-neighbours ranging from 20 to 30, with the best balanced-accuracy of $\sim$0.926$\pm$0.005, for n-neighbours of 26.

\textbf{Gaussian Naive Bayes (GNB):} For the GNB algorithm, we scan over the prior-probability parameter, which represents the initial model assumption for the LSP fraction in the ML-Blazar sample. A change in prior-probability affects the model's prediction power, as shown in Fig.~\ref{fig:GNB_Prior_Prob}. In this case, the algorithm forces a more significant weight to a given class when the assigned prior is larger than the real fraction. The balanced-accuracy is optimal when setting a prior-probability similar to the real fraction of LSP in the ML sample (of $\sim 0.70$). We test the situation where the training sample is balanced via SMOTE and see no improvement for the bal-acc. Therefore, the results for this algorithm goes with no balancing option and no fine-tuning (with prior-probability set to `auto').

In Table \ref{tab:model-metrics}, we summarize the metrics at optimal model conditions, that maximise the balanced-accuracy. All metric values correspond to the mean as calculated from 300 re-samples, with its associated $\sigma$. Fig.~\ref{fig:bal-acc-histo} shows the distribution of balanced-accuracy at the optimal setup. The models are highly effective in performing the classification task with relatively small $\sigma$, which points to the low level of noise associate both to the input features and the target variables \citep{Non-Case-Based-Interpretation-Caruana1999,Murdock2001}.

\subsection{Feature Importance}
\label{sec:feature-importance}

\begin{figure}
    \centering
    \includegraphics[width=0.9\linewidth]{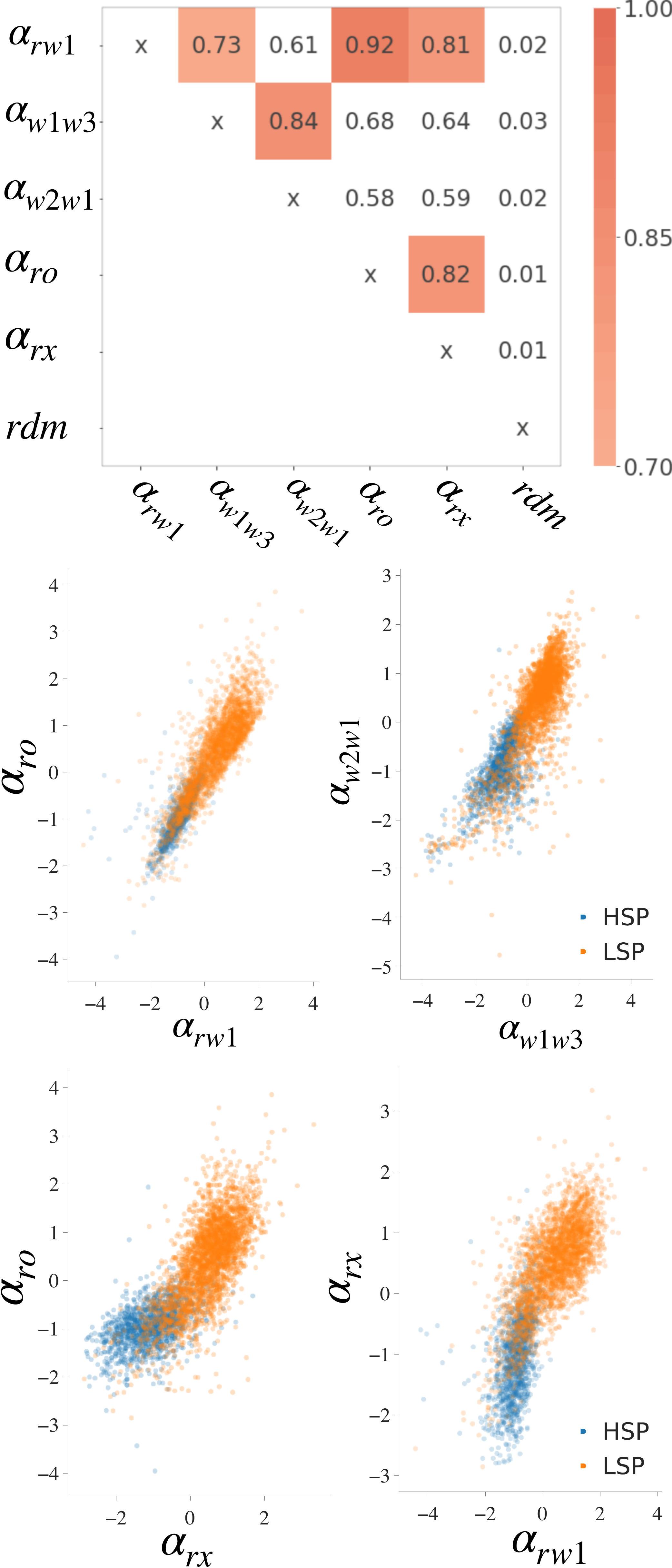}
    \caption{Top: Correlation matrix showing Pearson's coefficients to evaluate the linear correlation between features. Bottom: The four most correlated features, separated according to class: HSP (blue) and LSP (orange).}
    \label{fig:features-correlation}
\end{figure}

\begin{figure*}
    \centering
    \includegraphics[width=1.0\linewidth]{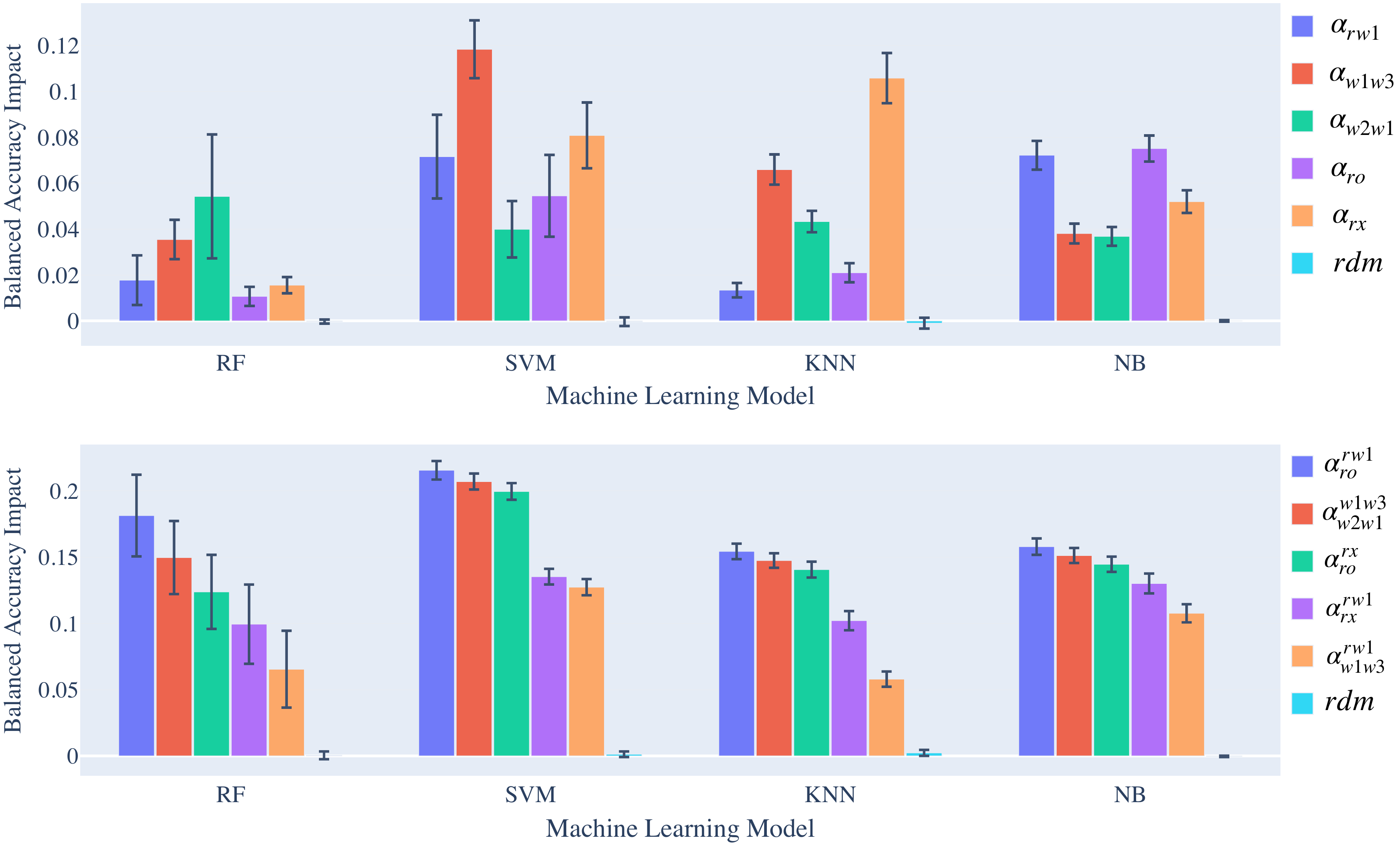}
    \caption{Top histogram: The impact of a single-feature permutation over the bal-acc. Bottom histogram: The absolute impact of joint-feature permutation over the bal-acc. The results refer to the machine learning models: RF, SVM, KNN and GNB. The points and error bars are the mean and standard deviation of each metric, considering 300 re-samples. Note, the random (rdm) feature is introduced by hand as a test-control and is composed by a standardized Gaussian noise: zero mean and unitary standard deviation.}
    \label{fig:feature-importance}
\end{figure*}

To explore the importance of each single-feature, one can try to visualise its impact on a given metric (e.g. balanced-accuracy) via a permutation test. However, this approach can only produce robust results in the case that all features are independent (or weakly correlated). If the data-frame has correlated features, an alternative is to apply the permutation jointly, to pairs of features, at once. Here we test both approaches and try to extract information related to the interpretation and consistency of our models.    

The permutation test consists in splitting the data-frame into train and test, for training an ML model as usual. For the test sample, however, the elements of each feature are shuffled or permuted. Consider `column i' to represent a given feature $\alpha_i$; The shuffling is done separately for each $\alpha_i$, and each time, the ML model estimates the class of each object in the permuted-test-sample$_i$. The absolute change in balanced-accuracy for each permuted-test-sample$_i$ is compared to the original balanced-accuracy of the test-sample \citep{Strobl2007}. In other words, if $\alpha_i$ is not correlated to other features, the information stored in each $\alpha_i$ is destroyed, and its direct impact on the balanced-accuracy metric is measured.

The permutation importance test does not depend on model parametrisation \citep[e.g. to interpret model coefficients,][]{breiman2001}, which renders it versatile to apply in all our cases. However, when features are correlated, e.g. features $\alpha_i$ and $\alpha_j$, the information about $\alpha_i$ is mapped into $\alpha_j$, and shuffling $\alpha_i$ alone is not enough to erase its information. As a result, the impact over the final balanced-accuracy is small and biased towards low importance. 

To visualise the relation between features, we build the correlation-matrix, Fig.~\ref{fig:features-correlation} (top), which shows the Pearson's coefficient between all combinations of spectral slopes ($\alpha$). We do add an extra column (rdm) to our data-frame, consisting of random Gaussian and standardized noise ( $\langle rdm \rangle \sim$0.0 and $\sigma$$\sim$1.0). The random column works as a test-control, given we do expect negligible impact over the balanced-accuracy when shuffling it. 

Indeed, our training data-frame has at least four pairs of spectral slopes with relevant correlation (Pearson's coefficient >\,0.8): $\alpha^{rw1}_{ro}$, $\alpha^{w1w3}_{w2w1}$, $\alpha^{rx}_{ro}$, $\alpha^{rw1}_{rx}$; and a weak one for $\alpha^{rw1}_{w1w3}$. We consider all those five pairs for a joint-feature permutation test. The scatter plots for the most relevant correlations are shown in Fig.~\ref{fig:features-correlation} (bottom).

We carried out the single-feature permutation test with the python library mlxtend \citep[MLX,][]{raschkas_2018_mlxtend} \textit{mlxtend.evaluate. feature-importance-permutation} \footnote{ \href{https://rasbt.github.io/mlxtend/}{The mlxtend:github documentation}.}. The MLX implementation has an internal option (num-rounds) which allows to randomly re-shuffle the feature permutation test n times (we adopt n=25) within a single loop, to probe the impact on balanced-accuracy. For the joint-feature permutation test we use the python library rfpimp\footnote{\href{https://github.com/parrt/random-forest-importances}{The rfpimp:github documentation}.} which implements the shuffling of pairs of features ($\alpha^{rw1}_{ro}$, $\alpha^{w1w3}_{w2w1}$, $\alpha^{rx}_{ro}$, $\alpha^{rw1}_{rx}$ and $\alpha^{rw1}_{w1w3}$). The results for both the single- and joint-features permutation tests are shown in Fig.~\ref{fig:feature-importance}.

To evaluate the degree of variability/uncertainty associated to each ML model, we run the entire test loop over 300 re-samples (for both the single and joint-feature permutation tests), and track the mean value and standard deviations of the impacts on the balanced-accuracy metric. The models are trained at their optimum, as described in Sec. \ref{sec:tunning}, and accounting for the unbalanced number of LSP \& HSP sources (either via considering a class-weight correction, or applying SMOTE over the training sample). We use the random split option to produce the train \& test samples (with 0.65 and 0.35 fraction, respectively) for each re-sample loop. 

It is interesting to compare the results for the single- and joint-feature tests. For the joint-feature test, the most relevant pair of features is $\alpha^{rw1}_{ro}$, for all models, followed by $\alpha^{w1w3}_{w2w1}$, $\alpha^{rx}_{ro}$, $\alpha^{rw1}_{rx}$ and $\alpha^{rw1}_{w1w3}$. For the single-feature test, each model seems -misleadingly- to have a strong dependence in a particular spectral slope. The RF, SVM, KNN and GNB models have respectively: $\alpha_{w2w1}$, $\alpha_{w1w3}$, $\alpha_{rx}$ and $\alpha_{ro}$ as the most important feature for each model, with no apparent reason behind the variability between models. However, the single-feature permutation test is biased because of the correlation between spectral slopes (as shown in Fig.~\ref{fig:features-correlation}), and the use of joint-features showed to be very useful to remove that bias.  

The RF model had the largest $\sigma$ for both single and joint-feature permutation tests. Note, this does not indicate the RF modelling is weaker (more unstable) concerning others, but that it is more sensible to the scrambling of a feature (or pair of features). In fact, the RF, SVM, KNN and GNB models are similar concerning model stability, as seen from the balanced-accuracy distribution in Fig.~\ref{fig:bal-acc-histo}, given that the $\sigma$ are similar (see Table \ref{tab:model-metrics}). 

The rdm feature (introduced by hand) showed nearly no impact over the balanced-accuracy during the single and joint-permutation tests. This was envisaged as a test-control, to which we expected null permutation impact.

\subsection{The intersection among classification models}
\label{sec:miss-rate} 

In addition to the performance of each algorithm alone, we also compared the classification of each object among the models. The approach was to generate 1000 randomly split train and test samples, to guarantee that all objects appear in the test set a reasonably large \textit{n} number of times (and smooth out fluctuations associated to \textit{n}). We used these 1000 samples to train all the ML algorithms, including Ludwig. 

Given the train and test process, we use two approaches to compare the classifications. The first one is to save the number of objects that are classified wrongly by at least one of the algorithms in each of the 1000 test samples, what we call as misclassification. The result of this approach is that a mean of 148.4 objects are misclassified at least once, with $\sigma$ of 19.2. Recall that our entire dataset has 4178 objects, so that the test sample has 1462 objects, meaning that about 10\% of them are classified wrongly by at least one algorithm. 
The second approach is to track and record how many times each object is misclassified, which we call by misclassification-rate (miss-rate). The result is that a total of 1244 objects have a miss-rate\,$\geq$\,1, meaning those are cases classified at least one time wrongly. The remaining 2934 objects were always classified correctly by all algorithms, suggesting that those have a clear distinction about which class they belong. 

\begin{figure}
    \centering
    \includegraphics[width=1.0\linewidth]{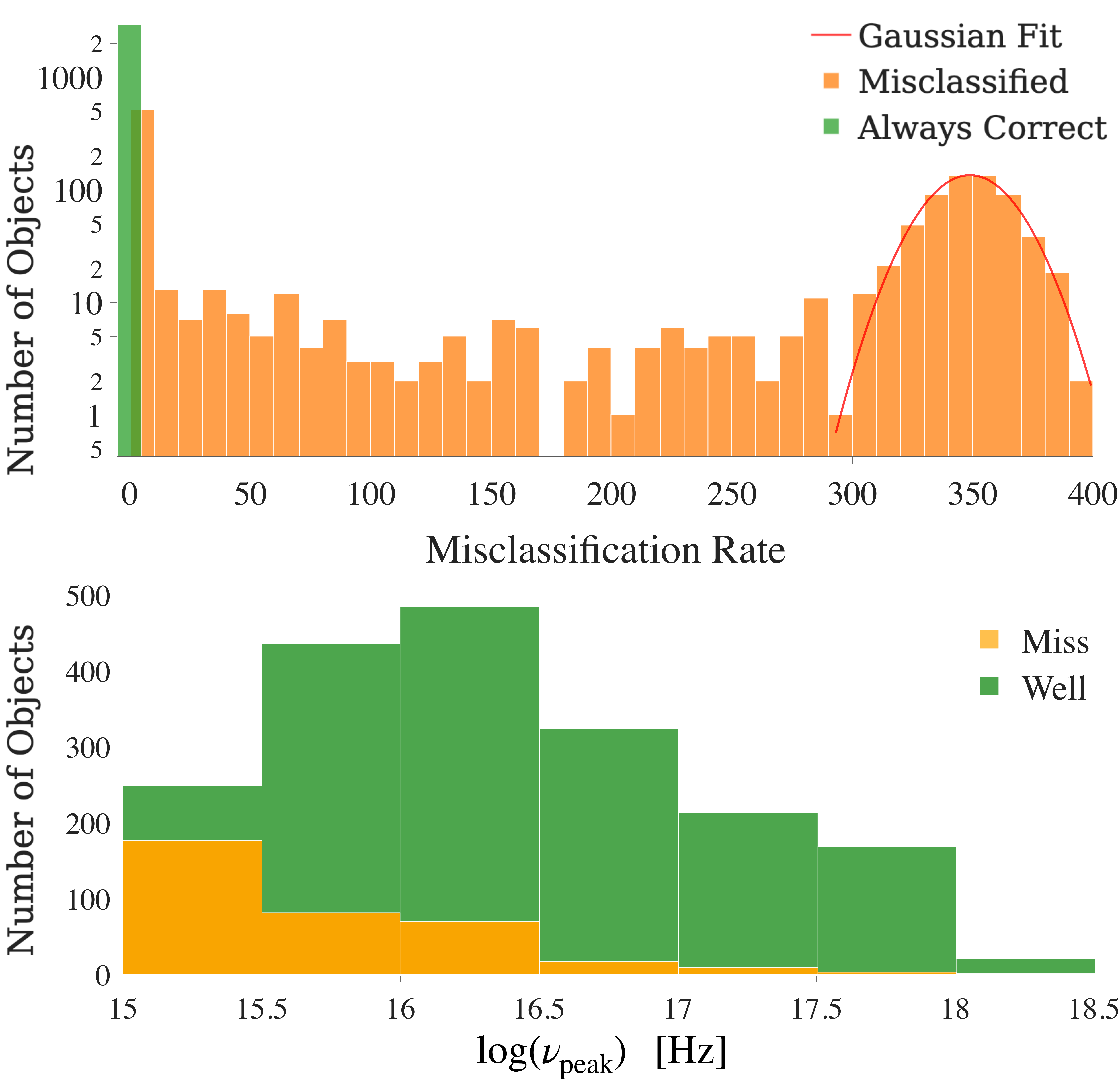}
    \caption{Top: Histogram of the miss-rate showing how many times an object is misclassified by at least one classifier, within 1000 re-samples. The vertical axes shows the number of objects in logarithmic scale. The orange bins represents the objects with miss-rate\,$\geq$\.1,  and the green bin represents the objects with miss-rate\,=\,0. The red line correspond to a Gaussian fit to the cases with miss-rate\,$\geq$\,262. Bottom: The log($\nu_{peak}$) [Hz] distribution associated to the most ambiguous cases with miss-rate\,$\geq$\,262 times (orange), and cases with miss-rate\,<\,262 (green). This histogram only considers the 1132 HSPs with confident $\nu_{peak}$ estimate, following the $\nu_{peak}$ flag from the 3HSP catalogue.}
    \label{fig:Misclassified}
\end{figure}

We plot the distribution of the 1244 objects with miss-rate\,$\geq$\,1, Fig.~\ref{fig:Misclassified} (top), highlighting the miss-rate = zero bin with 2934 objects. As seen, there is a group with hundreds of objects accumulating around the miss-rate of $\sim$350 counts, which follows approximately a Gaussian distribution having a mean of $348.7 \pm 0.7$ and $\sigma$ of $17.1\pm0.6$. We interpret this group of objects has an intrinsic ambiguity concerning the class they belong. Using a criteria of 5$\sigma$, we assume a threshold at a miss-rate $\geq$262 to mark (with a flag) these objects. Given the 1279 HSPs and the 2899 LSPs, 1073 (83.9\%) HSPs and 2499 (86.2\%) LSPs fall in the category with stable classification; 206 HSPs and 400 LSPs got the ambiguity-flag. The flag will be useful for future works, to help refine the classification and better understand the ML-Blazar data-frame.  

Fig.~\ref{fig:Misclassified} (bottom) shows a histogram according to log($\nu_{peak}$), separating between sources which are well classified most of times (miss-rate\,<\,262) and cases which are relatively harder to classify (miss-rate$\geq$262). The histogram only considers HSPs because the log($\nu_{peak}$) is only available for the entire HSP subsample \citep[from,][]{Chang2019}. The 5BZcat has no log($\nu_{peak}$) information. 



We notice a growing number of misclassified objects when the synchrotron peak approaches 10$^{15}$ Hz, revealing that the ML algorithms have increasing difficulty in classifying sources close to the decision border. In case of a human-based classification, both source variability (characteristic of blazars) and poor data coverage (for faint objects) can influence the determination of the $\nu_{peak}$ \citep[][]{Arsioli2015}. Those uncertainties may also drive the misclassification in ML-based methods. 

However, the information stored in the data-frame is not blind to variability. It builds over extensive work done with the 1WHSP, 2WHSP and 3HSP catalogues, selecting HSP sources while considering all available multifrequency \& multi-epoch data to compute the average $\nu_{peak}$. In this sense, both ML models and humans tend to misclassify border-sources, and it seems there would be no induced (or uncontrolled) cost by using ML instead of human classification, as discussed in \cite{Myths-ML-Interpretation-Zachary2016}. The feasibility of using ML instead of human classification will rely on comparing their uncertainties, along with other aspects as availability of computation power, run-time, and data/storage size.

\section{Conclusions}

We build the ML-Blazar data-frame condensing multifrequency information on Radio, IR, Optical, and X-rays, for sources listed in the 5BZcat and 3HSP catalogues. The ML-Blazar sample includes 4178 objects labelled as 1279 HSPs and 2899 LSPs. Section \ref{sec:selection} describes the cross-matching between catalogues and give details about the six flux channels available to describe each source.

The multifrequency fluxes were used to calculate spectral slopes ($\alpha$) that work as input features for training ML algorithms. In Sec. \ref{sec:feature-selection} we use statistical tests (KS and t-test) to compare the HSP and LSP distributions concerning each feature and select the five spectral slopes which better separate between HSPs and LSPs: $\rm \alpha_{rw_1}$, $\rm \alpha_{ro}$, $\rm \alpha_{rx}$, $\rm \alpha_{w_2 w_1}$, $\rm \alpha_{w_1 w_3}$. 

In Sec. \ref{sec:MLalgos}, we described five machine learning algorithms and the evaluation metrics used to train and test our blazar classification models. Those algorithms include Support Vector Machine (SVM), Random Forest (RF), K-Nearest Neighbours (KNN), Gaussian Naive Bayes (GNB), and Ludwig (auto-ML framework from Uber). Given that the ML-Blazar is unbalanced, we select the balanced-accuracy metric as a proxy for model optimisation. Besides, the training always corrects for unbalance via SMOTE-oversampling the HSPs, or by assigning larger weight to them.   

In Sec. \ref{sec:train-test} we study the influence of sample-size and test-fraction over the balanced-accuracy and shown that a 0.35 test-fraction can deliver low levels of uncertainly ($\sim$1\%) for the evaluation metrics, with no substantial penalty over the model's predictive power. In Sec. \ref{sec:tunning}, we describe the optimisation and best results for the SVM, RF, KNN, GNB, and Ludwig algorithms. We show that scanning over the parameter space available for each ML algorithm is an effective optimisations strategy. All data-points derived during the optimisation phase are mean values calculated from 300 re-sample, from which we extract the $\sigma$ as a measure of the uncertainty associated with the means.    
We test the importance of the input features for the SVM, RF, KNN and GNB algorithms, via a joint-feature permutation test, that account for the correlation between spectral slopes. The pair of features that produce the main impacts in prediction power is $\alpha^{rw1}_{ro}$, for all models, followed by $\alpha^{w1w3}_{w2w1}$, $\alpha^{rx}_{ro}$, $\alpha^{rw1}_{rx}$ and $\alpha^{rw1}_{w1w3}$. We have shown that supervised ML can be used to classify blazars  into HSPs and LSPs based on multifrequency information, reaching up to $\sim$93\% of balanced-accuracy with the SVM algorithm. We had a focus in optimisation strategies (to improve the balanced-accuracy) but also intended to provide the basis to investigate automated methods that can determine broadband spectral parameters, as the synchrotron $\nu$-peak.

The high balanced-accuracy ($\sim$93\%) shows that LSPs and HSPs are highly separable in the multidimensional space of spectral slopes. The separability could be connected to a physical distinction between the synchrotron-emission regime in each class. As known, in single-zone self-synchrotron Compton (SSC) scenarios, the transition between Thomson to Klein-Nishina emission regime happens at log($\nu_{peak}$)$\sim$14.7 [Hz] \citep[see the `Tramacere Plane',][]{gamma-region-Arsioli2018}, close to the LSP-HSP threshold at log($\nu_{peak}$)\,=\,15.0. The presence of external photon fields interacting with the blazar's jet (in External Compton scenarios, EC) can also play a whole for the interpretation. Moreover, \cite{Strong-Weak-Jets-Eileen2020} reports on the AGNs weak-jet and strong-jet modes (respectively, in inefficient and efficient accretion systems) that is tightly related to a distinction between LSPs and HSPs. The separability of LSPs and HSPs can be investigated in future, using unsupervised clustering methods with a focus on the physical difference between classes.


A multifrequency ML approach can be used to unveil new blazars and other types of sources, looking forward to explore the full potential of currently available astrophysical archives. We envision the application of ML algorithms as a highly effective tool to automate the extraction of spectral parameters and to classify astrophysical sources, especially useful for the upcoming generation of deep-sky surveys.

\section{Data Availability}

The data underlying this article are available in VizieR, at https://dx.doi.org/10.26093/cds/vizier. The datasets were derived from sources in the public domain. The main catalogues are the Radio: FIRST, NVSS, SUMSS, PMN and TAPMN \citep{Helfand2015,Condon1998,Manch2003,PMN-radio,ATPMN-radio}; IR: AllWISE survey  \citep{Cutri2013}; Optical: GAIA DR2, SDSS DR12, PanStars DR1, and Usno B.1 \citep{GAIA-DR2,SDSS-DR12,Chambers2016,Monet2003}; X-ray: 3XMM DR8, Swift-1SWXRT, Swift-XRTGRB , XMM-SL2, 2RXS-RASS, and Chandra V1.1. \citep{3XMM-DR8,SWIFT-XRT1,SWIFT-deep,XMMSL1,2RXS-RASS,Chandra}; Gamma-ray: 4FGL and 2BIGB \citep{4FGL,2BIGB}; Blazars: 5BZcat and 3HSP \citep{Massaro2009,Massaro2015,Chang2019}.

\section*{Acknowledgements}
BA is supported by S\~ao Paulo Research Foundation (FAPESP) with grant n. 2017/00517-4. PD is supported by S\~ao Paulo Research Foundation (FAPESP) with grant n. 2019/08956-2. We made use of archival data and bibliographic information obtained from the NASA/IPAC Extragalactic Database (NED), data and software facilities from the SSDC and Opne Universe Portals managed by the Italian Space Agency (ASI) and collaborators. We acknowledge the use of the TOPCAT and STILTS software packages (written by Mark Taylor, University of Bristol).




\bibliographystyle{mnras}
\bibliography{MLblazar} 








\bsp	
\label{lastpage}
\end{document}